\documentclass[referee,a4paper,12pt,traditabstract]{swsc} 

\usepackage{graphicx, graphics}
\usepackage{txfonts}
\usepackage[caption=false,font=footnotesize]{subfig}
\usepackage{rotating}
\usepackage{pdflscape}
\usepackage{epstopdf}
\usepackage{lineno}
\usepackage[authoryear,round]{natbib}
\usepackage[backref]{hyperref}
\usepackage{url}
\usepackage{multirow}

\hypersetup{colorlinks=true,citecolor=cyan,urlcolor=cyan,linkcolor=blue}

\begin{document}


   \title{Implementation and validation of the GEANT4/AtRIS code to model the radiation environment at Mars}

   
   \titlerunning{The AtRIS application to Mars}

   \authorrunning{Guo, Banjac and R\"{o}stel  et al.}

   \author{Jingnan  Guo
          \inst{1}
          \and
          Sa$\check{\rm{s}}$a~Banjac
          \inst{1}
          \and
          Lennart R\"{o}stel
          \inst{1}
          \and
          Jan C. Terasa
          \inst{1}           
          \and
          Konstantin Herbst
          \inst{1}
          \and
          Bernd Heber
          \inst{1}
          \and
          Robert F. Wimmer-Schweingruber
          \inst{1}
          }

   \institute{Institute of Experimental and Applied Physics (IEAP), University of Kiel,
              Leibnitzstr. 11, 24118 Kiel, Germany.
              \email{\href{mailto:guo@physik.uni-kiel.de}{guo@physik.uni-kiel.de}
            }}


 
  \abstract
{A new GEANT4 particle transport model -- the Atmospheric Radiation Interaction Simulator (AtRIS, Banjac et al., 2018) -- has been recently developed in order to model the interaction of radiation with planets. 
The upcoming instrumentational advancements in the exoplanetary science, in particular transit spectroscopy capabilities of missions like JWST and E-ELT, have motivated the development of a particle transport code with a focus on providing the necessary flexibility in planet specification (atmosphere and soil geometry and composition, tidal locking, oceans, clouds, etc.) for the modeling of radiation environment for exoplanets. Since there are no factors limiting the applicability of AtRIS to Mars and Venus, AtRIS' unique flexibility opens possibilities for new studies.

Following the successful validation against Earth measurements (Banjac et al., 2018), this work applies AtRIS with a specific implementation of the Martian atmospheric and regolith structure to model the radiation environment at Mars. 
We benchmark these first modeling results based on different GEANT4 physics lists with the energetic particle spectra recently measured by the Radiation Assessment Detector (RAD) on the surface of Mars.
The good agreement between AtRIS and the actual measurement provides one of the first and sound validations of AtRIS and the preferred physics list which could be recommended for predicting the radiation field of other conceivable (exo)planets with an atmospheric environment similar to Mars.}

\keywords{ Particle Radiation in Space, Particle transport model, Martian Exploration, Planetary Space Weather }

\maketitle

\section{Introduction}\label{sec:intro}
There are mainly two types of energetic particles in the heliosphere that may impose risks for deep space and planetary missions: galactic cosmic rays (GCRs) and solar energetic particles (SEPs). 
GCRs are energetic charged particles, comprised of 2\% electrons and 98\% atomic nuclei with the later contributed by 87\% protons, 12\% helium, and about 1\% heavier nuclei (Z$\ge$ 3) \citep{simpson1983}.
They have energies from less than 1 MeV/nuc up to hundreds of TeVs with a power-law distribution at energies above $\sim$ GeV/nuc \citep{allkofer1975introduction}. 
SEPs on the other hand are energetic particles (mainly protons and electrons) emitted from the Sun and accelerated by solar flares and/or Coronal Mass Ejection (CME) associated shocks.
SEPs generally have energies from a few keV up to several hundreds MeV (occasionally even reaching 1 or 2 GeV or further above) and can reach significantly higher fluxes at these energies compared to background GCRs. 

It is likely that the first human-visited planet will be our neighbour planet Mars. It has a thin atmosphere with its surface pressure less than 1 percent of that at Earth's surface making it much easier for high energy particles to reach the Martian surface. 
Therefore, the assessment of the Martian radiation environment is necessary and fundamental for (a) mitigating radiation risks for near-future robotic and crewed missions and (b) better understanding the impact of energetic particles on the preservation of organic biosignatures on Mars. 
Primary GCRs and SEPs passing through the Martian atmosphere may undergo inelastic interactions with the ambient atomic nuclei losing their energies and also creating secondary particles via spallation and fragmentation processes.  
These secondary particles may further interact with the atmosphere as they propagate downwards and even with the Martian regolith, finally resulting in very complex spectra including both primaries and secondaries at the surface of Mars \citep[e.g.,][]{saganti2002}. 

There are various particle transport codes such as HZETRN \citep[High charge (Z) and Energy TRaNsport,][]{slaba2016solar, wilson2016}, PHITS \citep[Particle and Heavy Ion Transport code System,][]{sato2013} and GEANT4 \citep[GEometry And Tracking,][]{agostinelli2003, allison2016recent} which can be employed for studying the particle spectra and radiation dose at Mars. 
Various studies have also combined these particle transport codes with different GCR and/or SEP spectra for estimating the radiation exposure on the surface of Mars \citep[e.g.,][]{simonsen1990radiation, simonsen1992mars, saganti2004radiation, keating2005model, deangelis2006modeling, mckenna2012characterization, ehresmann2011, guo2017dependence}.
Articles collected in a special issue at Life Sciences in Space Research \citep{HASSLER20171} have included most recent studies to model the radiation environment at Mars which is also compared with in situ measurement at the surface of Mars \citep{EHRESMANN20173, guo2017neutron} by the radiation assessment detector \citep[RAD,][]{hassler2012} on board the Curiosity rover belonging to the Mars Science Laboratory \citep[MSL,][]{grotzinger2012mars}. 

In particular, PLANETOCOSMICS (http://cosray.unibe.ch/$\sim$laurent/planetocosmics/) is a toolkit based on GEANT4 with a specific application purpose to simulate particle transport in planetary magnetic fields as well as interactions when passing through planetary environments and creating secondary particles \citep{desorgher2006planetocosmics}.
Modeling the radiation environment on the surface of Mars using PLANETOCOSMICS has been carried out by previous researchers \citep[e.g., ][]{dartnell2007modelling, gronoff2015computation, matthia2016martian, ehresmann2011, guo2018generalized} and has been validated by \citet{matthia2016martian} when compared to energetic charged and neutral particle spectra on the surface of Mars measured by MSL/RAD.

Since the landing of MSL at the Gale Crater of Mars on August 6, 2012, RAD has been providing the first in situ detection of the radiation environment at the surface of Mars \citep{hassler2014}. 
These novel measurements provide evaluations of the Martian radiation level at an atmospheric depth around 22 g/cm$^2$. 
It has been found that the measured GCR radiation dose rate is anti-correlated with the surface pressure (proportional to the atmospheric column depth) which changes both daily and seasonally \citep{rafkin2014, guo2015modeling, guo2017dependence} up to $\pm$ 25\%.
Meanwhile the varying heliopsheric conditions modulate the GCR fluxes and thus RAD measured radiation is anti-correlated with solar activity both in the long term evolution \citep{guo2015modeling} and in the short term due to e.g., interplanetary CMEs and their associated shocks passing Mars \citep{witasse2017interplanetary, guo2018measurements, von2018using, winslow2018opening}. 
RAD also measures the flux spectra of energetic charged particles, such as protons, deuterons, tritons, $^3$He and $^4$He, carbon, nitrogen, oxygen and iron ions which penetrate downwards and stop within the detector set with energies up to about 100 MeV/nuc \citep{ehresmann2014, EHRESMANN20173}. 
With an inversion method exploiting the response matrix of the neutral particle detection efficiency in the scintillators, RAD can also measure spectra of neutral particles on Mars from $\sim$ 10 to 900 MeV \citep{koehler2014, guo2017neutron}. 

The last solar cycle (no. 24) has been rather quiet and RAD has only detected a few SEP events at the Martian surface and most of them are rather insignificant apart from the September 10, 2017 event \citep{zeitlin2018analysis, ehresmann2018energetic, guo2018modeling}.
This was the first ground level enhancement (GLE) resulting from a very intense SEP detected at the surface of two different planets: Earth and Mars.
In general, SEP events are sporadic, often rather impulsive and could be extremely hazardous especially when the observer has a direct magnetic connection to the particle acceleration region and injection site at the Sun. 
In the case of Mars, there is no global magnetosphere that could shield the atmosphere from energetic particles. In addition, its thin atmosphere can only stop charged particles below $\sim$ 150 MeV/nuc \citep{guo2018generalized}.
Therefore it is important to reliably model SEP events and their induced radiation in order to give immediate and precise alerts for future human missions to Mars. 
To do so, it is essential to well understand the interactions of atmospheric molecules with incoming particles. 

Recently, a new GEANT4 particle transport model -- the Atmospheric Radiation Interaction Simulator (AtRIS, Section \ref{sec:model_atris}) -- has been developed by \citet{banjac2018} in order to model the interaction of radiation with planets. 
The upcoming instrumentational advancements in the exoplanetary science, in particular transit spectroscopy capabilities of missions like JWST and E-ELT, have motivated the development of this particle transport code with a focus on providing the necessary flexibility in planet specification (atmosphere and soil geometry and composition, tidal locking, oceans, clouds, etc.) for the modeling of exoplanets. 
The application of AtRIS to model the radiation environment on Mars has been realized and validated for the first time in this study. 

\section{Model Description and Implementation}\label{sec:model_description}

\subsection{The Mars Climate Database} \label{sec:model_mcd}

The Mars Climate Database (MCD, http://www-mars.lmd.jussieu.fr) offers the possibility to access Martian atmospheric properties, such as temperature, density and composition, for different altitudes, seasons and even the time of the day on Mars.
MCD has been developed using different Martian atmospheric circulation models which are further compared and modified by the observation results from past and current Mars missions \citep{lewis1999climate}. 
Therefore, it provides a Martian atmospheric environment which can be implemented into the planetary particle transport toolkit for the purpose of simulating high energetic particles interacting with the Martian atmospheric atoms. 

\begin{figure}
\centering
\includegraphics[scale=0.5]{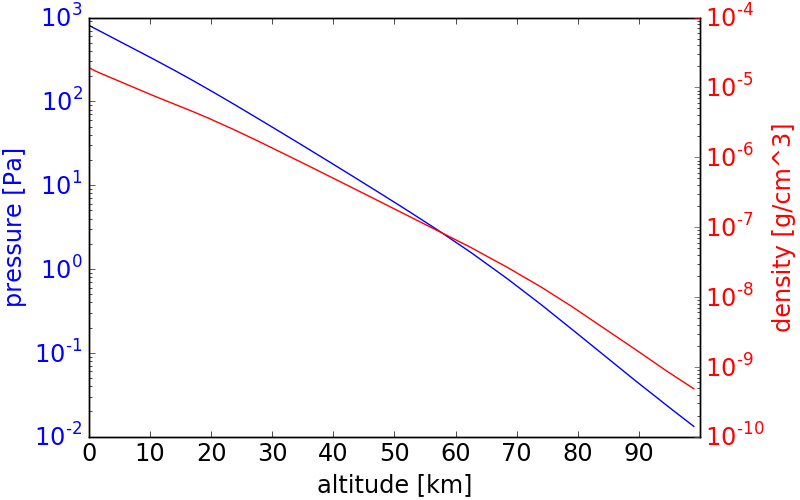}
  \caption{Martian atmospheric structure at Gale Crater implemented in the AtRIS simulations as imported from the MCD model.}
     \label{fig:atmosphere}
\end{figure}

In our atmospheric setup for the AtRIS particle transport model, we use the composition, density and temperature profiles from MCD at Gale Crater on Mars where  MSL's rover Curiosity landed on August 6, 2012 (coordinate: 4.5$^\circ$S, 137.4$^\circ$E). 
The {elemental} composition of the Martian atmosphere consists of C, O, N, Ar, and H with more than 95\% of the molecules being CO$_2$. 
The atmospheric condition is set to be "clim aveEUV" for climatology scenario with average solar EUV radiation. 
The Martian solar longitude is set to be 200.5$^{\circ}$ when the surface pressure is close to the annual average pressure at Gale Crater measured by MSL $\sim$ 840 Pa \citep[e.g.,][]{guo2015modeling}. 
Figure \ref{fig:atmosphere} shows the atmospheric structure, i.e, density and pressure versus the atmospheric altitude used for the AtRIS simulations. 

\subsection{Model description: AtRIS}\label{sec:model_atris}

GEANT4 is a Monte Carlo approach widely used for simulating the interactions of particles as they traverse matter \citep{agostinelli2003}. 

The Atmospheric Radiation Interaction Simulator, AtRIS, is a GEANT4 based particle transport code developed to simulate the propagation of energetic particles through planetary atmosphere and regolith. 
AtRIS allows rather flexible geometry and composition definitions of the planet. 
This flexibility and ease of use has been achieved through a custom interface called the \textit{planet specification format}, that has been optimized for the specification of planets and their atmospheres. 
For Earth, an interface based on the so-called NRLMSISE-00 \citep{picone2002nrlmsise} model is provided within AtRIS.
Similarly, for Mars, the Mars Climate Database (MCD) interface has been implemented (see Section~\ref{sec:model_mcd}). 

AtRIS can calculate ion and electron pair production rates, secondary particle distributions (as a function of energy, directionality, planet altitude and so on), as well as absorbed and equivalent dose rates for a 30 cm diameter ICRU sphere phantom composed of water.
The tracking of charged particles through magnetic fields is not implemented in AtRIS.
As Mars lacks a global magnetic field, it serves as a good validation object for particle propagation modeling of AtRIS. 
Control over hadronic and electromagnetic processes is provided via the standard GEANT4 messenger and compounded physics list naming scheme \citep{allison2016recent, collaboration2017geant4}. Different physics lists tested and compared in this study will be discussed in detail in Section \ref{sec:validation_matrix}.
 
The main features of AtRIS are i) the planet specification format as explained above, ii) the atmospheric response matrices (ARMs), quantifying the relation between primary energy and altitude-dependent ionization as well as equivalent/absorbed dose rate, and finally, iii) the spectrum folding procedure used to calculate net quantities like the electron-ion pair production rate, by implementing a convolution  of a measured spectrum and the ARM. 
A more detailed description is given in  \citep{banjac2018}, where the results of AtRIS were compared with and validated against different kinds of Earth measurements (i.e., ion pair production and secondary particle fluxes, absorbed dose rate and dose equivalent).

In the current AtRIS setup for the Martian environment as shown in Table \ref{table:models}, we use a {sphere} with a radius of 3390 km (approximately the average radius of Mars) representing Mars. The soil composition is approximated as 50\%Si, 40\%O, and 10\% Fe (mass fractions) and the crust (soil) sheet is 100~m thick. 
The maximum height of the atmosphere from the MCD model is 100~km which is divided into 500~m thick layers. The accumulated column depth at the surface is about 22 g/cm$^2$ corresponding to a surface pressure of 830 Pa (Figure \ref{fig:atmosphere}). Table \ref{table:models} provides the main features of the atmospheric and regolith structures used for the AtRIS simulations. 

\subsection{The matrix realization of AtRIS}\label{sec:model_matrix}

Recently, \citet{guo2018generalized} have developed a generalized approach based on the GEANT4/PLANETOCOSMICS transport code and the MCD Martian environment setups to quickly model the Martian surface radiation level of any given incoming proton/helium ion spectra. 
Such an approach called "Planetomatrix" can be mathematically described by atmospheric response matrices and visually illustrated as 2-d histograms (Figure 1 and 2 of \citet{guo2018generalized}). 

Similar to this Planetomatrix approach, we construct the "response function" of the Martian atmosphere based on simulated particle spectra from AtRIS. 
We can describe the statistical transformation of the atomic and nuclear interaction process for a particle spectrum (of particle type i as a function of energy $E_0$) above the Martian atmosphere resulting in a particle spectrum (of particle type j as a function of energy $E$) on the Martian surface in a matrix $\rm{\bar M_{ij}(E_0, E)}$. 
As particles could also interact with the Martian atmosphere and regolith and produce albedo particles contributing to the upward fluxes, we have also considered the generation of such upward particles in our simulations.  
Since the energy spectra of upward- and downward-traveling particles are dissimilar, we have separately constructed the upward and downward directed matrices for each primary-secondary case. 
Once such a matrix is constructed, it can be folded with any incoming SEP or GCR spectra within the energy range of the simulated particles for calculating the surface spectra without re-running the particle transport simulations. 
The upward and downward secondary particle spectra generated by different types of primary particles can be combined. 
The following steps are used for constructing the matrix and folding it with a given incoming spectra to obtain the surface spectra:
\begin{enumerate}
    \item Simulate primary particles (type $i$) with energy in the range of 1 and 10$^5$ MeV through the Mars AtRIS model described in Section \ref{sec:model_atris}. A flat spectrum over 50 energy bins distributed uniformly in logarithmic scale is used. There are $N_{i}(E_0)$ particles simulated in each energy bin. 
    \item Based on the simulation results create a matrix $\rm{M_{ij}(E_0, E_j)}$ for each secondary particle type $j$ with certain directions, e.g., downward-directed protons on the surface of Mars generated by primary protons. 
    Each column of the matrix is a histogram of secondary particles $H_{i}(E_j)$ representing the number of particles at energy $E_j$ created by $N_{i}(E_0)$ primary particles with energy $E_0$. 
    \item Divide each histogram $H_{i}(E_j)$ in $\rm{M_{ij}(E_0, E_j)}$ by the number of simulated particles located in each incoming energy bin $N_{i}(E_0)$ to generate the normalized histogram due to a single primary particle with energy $E_0$. The normalized histograms constitute the normalized matrix which is $\rm{M_{ij}^N(E_0, E_j)} \coloneqq \rm{M_{ij}(E_0, E_j)}/N_{i}(E_0)$.
    \item For a given GCR/SEP spectrum $f_{i}$ which is often an isotropic flux in deep space (e.g., with units of [particles/sec/m$^2$/sr/MeV]), first interpolate this spectrum using the incoming energies ($E_0$) of the matrix $\rm{M_{ij}(E_0, E_j)}$ so that $f_{i} = f_{i}(E_0)$. 
    Then multiply it with $W(E_0)$ which is the incoming bin width [MeV] of the matrix such that $F_{i}(E_0) = W(E_0) f_{i}(E_0)$ with the unit of [particles/sec/m$^2$/sr].
    \item Calculate the total primary particle count rate [particles/sec] taking into account of the integrated geometric factor of particles arriving at the planet by multiplying the above $F_{i}(E_0)$ with $\Omega_{in} A_{in}$. Here $A_{in}$ is the area of which primary particles were fed into the simulation. 
    When using a sphere covering the top of the planetary atmosphere as the source sphere, $A_{in}$ is 4$\pi R^2$ with $R$ [meter] being the radius of this sphere (i.e., $R_{top}$ in Table \ref{table:models}). 
    $\Omega_{in}$ is the integration of the cosine of zenith angle of the source particles and is $\pi$ when they were injected isotropically inward from the source sphere. 
    \item  Fold such scaled incoming particle count rate (energy dependent) with the normalized matrix to obtain the histogram of secondaries [particles/sec] corresponding to each column of the matrix (at certain incoming primary energy). 
    \item  Scale the secondary particle histogram at each column into a differential spectra per geometric factor by dividing it with $\Omega_{out} A_{out} W(E_j)$. Here $W(E_j)$ is the energy bin width [MeV] of the secondary particle histogram. $A_{out}$ is the surface area of the planet where the output secondaries are counted and it is 4$\pi R_{surf}^2$ with $R_{surf}$ (see Table \ref{table:models}) in centimeter. And the $\Omega_{out}$ is the integration of the cosine of zenith angle of secondaries crossing the surface plane.
\end{enumerate}

In this study, we consider three cases of secondary particle directions from AtRIS simulations: a) all secondaries propagating downwards with the flux-weighted solid angle $\Omega_{out}$ of $\pi$, b) all secondaries propagating upwards with $\Omega_{out} = \pi$ and c) secondary downward directed particles with a zenith angle within 36$^\circ$, i.e, $\Omega_{out} = 1.1$ sr. 
This corresponds to the inner cone angle used for counting downward propagating charged particles stopping inside the RAD detector \citep{hassler2012}.

Finally the surface secondary spectra ${F}_{ij}(E_j)$ induced by primary particle spectrum $f_i(E_0)$ obtained through step 1-7 has unit of [particles/sec/cm$^2$/sr/MeV].
The above procedure can be also mathematically summarized as:
\begin{eqnarray}\label{eq:matrix_derive}
{\rm{\bar M}}_{ij}(E_0, E_j) &=& \frac{ {\frac{{\rm{M}}_{ij}(E_0, E_j)}{N_{i}(E_0)}} {\Omega_{in} A_{in}} } {\Omega_{out} A_{out} W(E_j)} \nonumber \\
&=& \frac{\Omega_{in} A_{in}}{\Omega_{out} A_{out}} \frac{\rm{M_{ij}^N(E_0, E_j)}}{W(E_j)},
\end{eqnarray} 
where $\rm{\bar M_{ij}(E_0, E_j)}$ is the atmospheric response matrix which takes into account the scaling of the energy bin widths of the output histograms and the geometric factors used in the simulation. 
The surface spectrum of particle type j resulting from primary particle type i is the multiplication of the matrix with the primary particle spectrum and this operation is essentially the sum product over the second axis ($E_0$) of $\rm{\bar M_{ij}}$ and $F_i$, i.e., 
\begin{eqnarray}\label{eq:flux_derive}
{F}_{ij}(E_j) &=& \int_{E^{k}_{0}}^{E^{k+1}_{0}} {\rm{\bar M_{ij}}(E_0, E_j)} f_{i}(E_0) W(E_0) dE_0 \nonumber \\
&=& \int_{E^{k}_{0}}^{E^{k+1}_{0}} {\rm{\bar M_{ij}}(E_0, E_j)} F_{i}(E_0) dE_0.
\end{eqnarray} 

The simulation is often set up with the areas $A_{in}$ and $A_{out}$ equal to each other where the planet is approximated as a rectangular box which has the same size on top of the atmosphere and on the regolith surface. 
In the current AtRIS/MCD setup, $A_{in}$ is only slightly bigger than $A_{out}$ as $R_{top}$ is the radius of the planet including the atmospheric layer and it is slightly larger than $R_{surf}$. 
Given three cases of solid angle of secondaries detected on the surface, i.e., downward, upward and within 36$^\circ$ zenith angle (RAD inner cone), a matrix for each case can be constructed for certain secondary particles induced by given primary particle type. 

Although the construction of each matrix is time-consuming, the multiplication of different input spectra $F_{i}(E_0)$ with such a matrix to generate different surface spectra ${F_{j}(E_j)}$ is very much simplified. 
The spectra of the surface secondary particle are the combined spectra resulting from different types of primary particles which arrive at the top the Martian atmosphere: 
\begin{eqnarray}\label{eq:matrix_all}
{F}_{j}^{dn}(E_j) &=& \sum_{i} \int_{E^{k}_{0}}^{E^{k+1}_{0}} {\rm{\bar M}}_{ij}^{dn}(E_0, E_j) F_{i}(E_0) dE_0\\
{F}_{j}^{up}(E_j) &=& \sum_{i} \int_{E^{k}_{0}}^{E^{k+1}_{0}} {\rm{\bar M}}_{ij}^{up}(E_0, E_j) F_{i}(E_0) dE_0\\
{F}_{j}^{RAD}(E_j) &=& \sum_{i} \int_{E^{k}_{0}}^{E^{k+1}_{0}} {\rm{\bar M}}_{ij}^{RAD}(E_0, E_j) F_{i}(E_0)dE_0.
\end{eqnarray} 
Here ${F}_{j}^{dn}(E_j)$, ${F}_{j}^{up}(E_j) $ and ${F}_{j}^{RAD}(E_j)$ represent the energy spectra of secondary particle type $j$, resulting from various primary particle types, averaged in the downward, upward and within 36$^\circ$ zenith angle respectively. 
As a first step of the model verification, we have simulated protons and $^4$He ions as primary particles because they constitute the majority of GCR particles. 
We considered the secondary charged particles (type j) whose spectra are also measured by RAD on the surface of Mars, such as protons, deuterons, tritons, $^4$He and $^3$He ions as output particles. 

\section{Validation of AtRIS}\label{sec:validation}

\begin{table}
	\centering
	\begin{tabular}{|r|c|c|c|c|}
		\hline
		\hline
		& model A & model B & model C & model D \\
		\hline
		physics lists  & QGSP\_BIC\_HP  & QGSP\_BERT\_HP & FTFP\_BERT\_HP & FTFP\_INCLXX\_HP\\
		\hline
		atmosphere & \multicolumn{4}{c|}{100 km thick spherical shell: $R_{surf}$ = 3390 km, $R_{top}$=3490 km}\\
		setup & \multicolumn{4}{c|}{MCD atmospheric profile at Gale Crater with a total column depth of $\sim$ 22 g/cm$^2$}\\
		\hline
		regolith & \multicolumn{4}{c|}{100 m thick of soil spherical shell beneath the atmosphere} \\
		setup & \multicolumn{4}{c|}{1.7 g/cm$^3$, 50\%Si, 40\%O, and 10\% Fe} \\
		\hline
	\end{tabular}
	\caption{Martian atmospheric and regolith properties used in AtRIS and four different physics lists of GEANT4 (more details in the text) tested in this study.}
	\label{table:models}
\end{table}

GEANT4 (version 10.4.p02 used here) offers a wide variety of models for handling physical processes within different energy ranges.
It is not yet entirely clear what is the most accurate and efficient physics list describing high energy (from hundreds of MeV to tens of GeV) particles integrating with the Martian atmosphere \citep{matthia2016martian}. 
One of the goals of MSL/RAD is to help validate the appropriate physics list which could precisely model the high energetic cosmic ray interaction with the Mars atmosphere \citep{hassler2012}. 

In this study, we employ four different physics lists (Table \ref{table:models}): QGSP\_BIC\_HP  (model A), QGSP\_BERT\_HP  (model B),  FTFP\_BERT\_HP  (model C) and also FTFP\_INCLXX\_HP (model D)  when applying AtRIS to model the Martian radiation environment. 
These physics lists are named based on the combination of various models in different energy ranges \citep{collaboration2017geant4}: 
\begin{itemize}
\item QGS stands for the Quark Gluon String model for high energy particles ($> \sim$ 20 GeV).
\item FTF represents the Fritiof model for particles $> \sim$ 5 GeV.
\item P stands for the Precompound model used for de-excitation process for nucleon-induced reactions below 1-2 MeV.
\item BIC uses the Binary Cascade Model for particles $< \sim$ 10 GeV.
\item BERT uses the Bertini Cascade Model in the middle energy range of $< \sim$ 10 GeV.
\item INCL represents the Li\'ege Intra-nuclear Cascade model in the middle energy range.
\item HP option switches on the high precision neutron elastic and inelastic scattering model for neutrons below 20 MeV.
\item At energies where different physics models are overlapping, a weighted combination of various models is considered.
\end{itemize}

Specifically speaking, model A) QGSP\_BIC\_HP and model B) QGSP\_BERT\_HP use the same Quark Gluon String (QGS) model for the high energy range while different cascade models for the lower energy range ($< \sim$ 10 GeV). 
FTFP\_BERT\_HP in model C is recommended by Geant4 collaboration "for cosmic ray applications where good treatment of very high energy particles is required" \citep{collaboration2017geant4}. 
It contains all standard electromagnetic (EM) processes by default. It uses Bertini-style cascade for hadrons $<$ 5 GeV and the FTF \citep{andersson1987model, nilsson1987interactions} model for simulating the interaction of mesons, nucleons and hyperons in the 3 GeV - 100 TeV energy range.	
Finally and in model D, the Li\`ege Intra-nuclear Cascade model INCL++ has been recently extended in GEANT4 to handle reactions between 3 and 15 GeV incident energies \citep{mancusi2014extension} and extensive benchmarks have shown the INCL++ model has a very good predictive power for the particles related to neutron production in spallation reactions \citep{leray2011results}. 
A map of physics models in different energy ranges used for the INCLXX physics list can be found here http://irfu.cea.fr/dphn/Spallation/physlist.html. 

We have validated the accuracy of the AtRIS approach when applied with different physics lists (named model A, B, C and D) to the Martian environment, especially for the generation of charged particle spectra by primary protons and $^4$He ions. 
We first compare the atmospheric response matrices of each primary-secondary pair when using different physics lists as shown in Section \ref{sec:validation_matrix}. 
Then we compare such modeled surface secondary spectra based on different physics lists with the MSL/RAD measurements as explained in Section \ref{sec:validation_RAD}. 

\subsection{Matrices constructed via AtRIS}\label{sec:validation_matrix}

Using the matrix approach described in Section \ref{sec:model_matrix}, we compare the atmospheric response matrices derived from above four different models: A) QGSP\_BIC\_HP, B) QGSP\_BERT\_HP, C) FTFP\_BERT\_HP and D) FTFP\_INCLXX\_HP. 
To ease the comparison between different simulation results, normalized matrices $\rm{M_{ij}^N(E_0, E_j)}$ (step 3 in Section \ref{sec:model_matrix}) have been plotted and discussed. 

The interaction of primary protons with the Martian atmosphere has been modeled and atmospheric response matrices have been constructed for five different secondary types including Hydrogen isotopes (proton, deuteron and triton) and Helium isotopes ($^3$He and $^4$He). Each secondary type is differentiated as downward- and upward-directed with the respective matrix. 
A total of ten matrices have been generated for primary proton interactions. Similarly ten other matrices represent the primary $^4$He ion interaction with the Martian atmosphere. 
Although there are other secondaries such as electrons, muons, positrons which also contribute to the radiation environment on Mars, we focus on the five hydrogen and helium isotopes as their spectra have been well measured on the surface by MSL/RAD \citep{ehresmann2014} and a direct model-observation comparison is possible. 

\subsubsection{Primary proton matrices}
\begin{figure}
	\centering
	\includegraphics[scale=0.21]{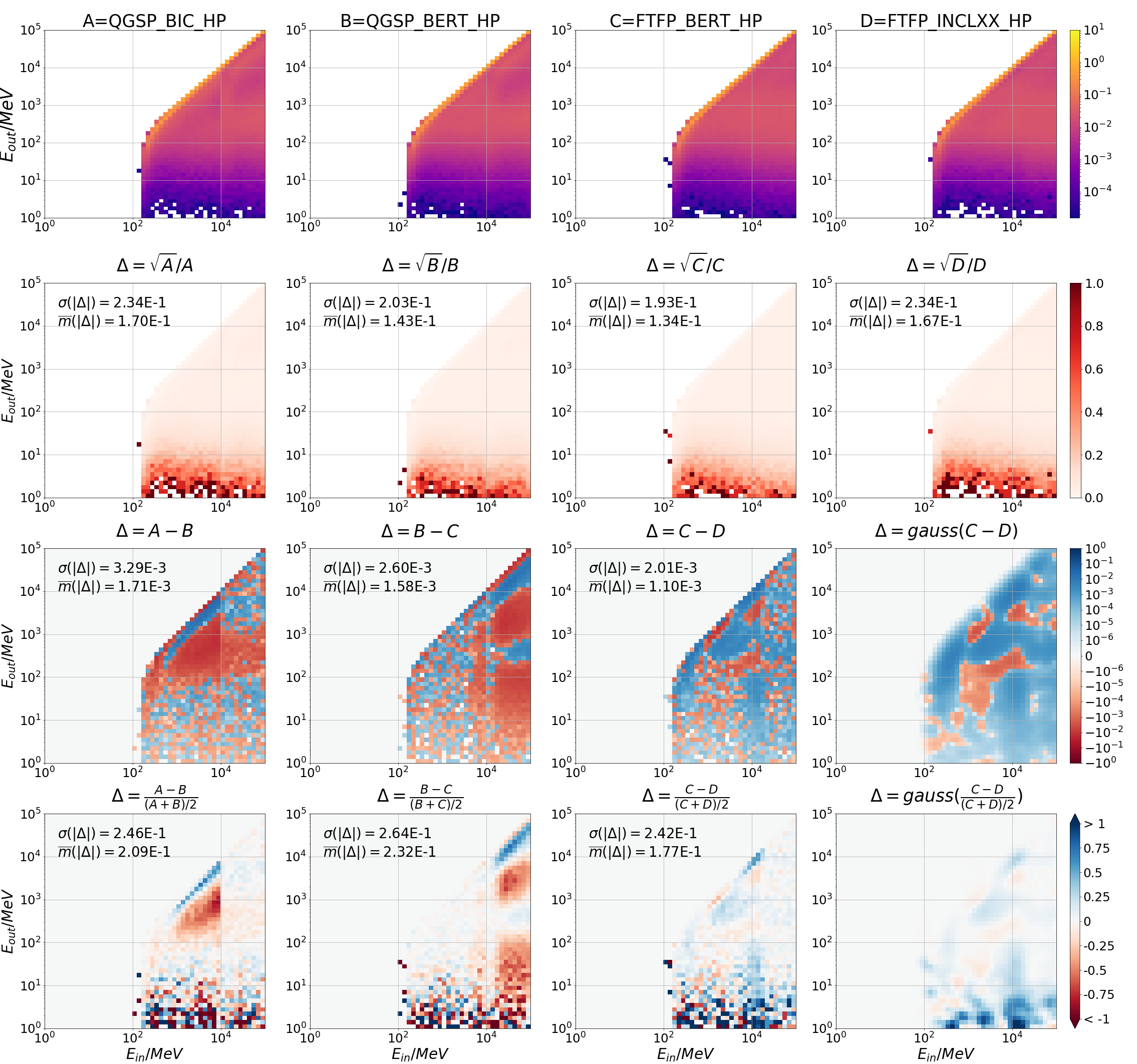}
	\caption{Matrices which describe the surface downward protons created by the primary protons. First row: the normalized Martian atmospheric matrices ($\rm{M_{ij}^N(E_0, E_j)}$) of primary protons generating surface downward directed protons from physics lists A, B, C and D (Table \ref{table:models}). 
	Second row: Normalized statistical uncertainty of the matrices. 
	Third row: Image Differencing Matrices (IDM) of model A and B (1st column),  B and C (2nd column), C and D (3rd column) which is also visualized with a 2D Gaussian filter applied (4th column). 
	Fourth row: the normalized IDM. 
	$\bar{m}$ represents the average value in each matrix (while bins with zero statistics have been excluded) and $\sigma$ is the standard deviation.
	All matrices are shown in the energy range from 1 MeV to 100 GeV (in logarithmic scale) for both input bins and output bins. 
	}
	\label{fig:matrix_p-p_dn}
\end{figure}

\begin{figure}
	\centering
	\includegraphics[scale=0.21]{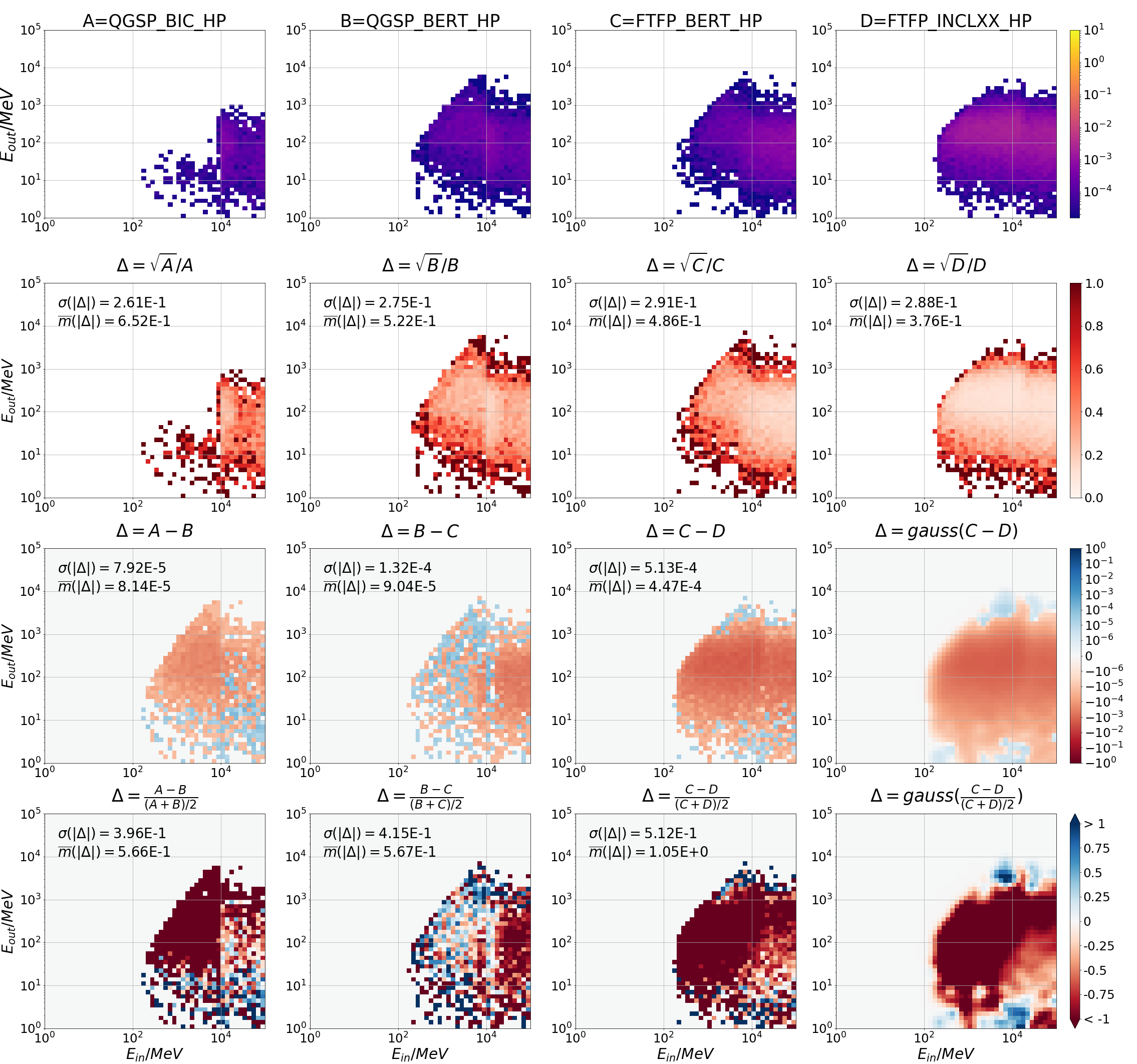}
	\caption{Matrices which describe the surface downward deuterons created by the primary protons. First row: the normalized Martian atmospheric matrices of primary protons generating surface downward directed $^2$H ions from physics lists A, B, C and D (Table \ref{table:models}). 
	Second row: Normalized statistical uncertainty of the matrices. 
	Third row: Image Differencing Matrices (IDM) of model A and B (1st column),  B and C (2nd column), C and D (3rd column) which is also visualized with a 2D Gaussian filter applied (4th column). 
	Fourth row: the normalized IDM. 
	$\bar{m}$ represents the average value in each matrix (while bins with zero statistics have been excluded)  and $\sigma$ is the standard deviation.
	All matrices are shown in the energy range from 1 MeV to 100 GeV (in logarithmic scale) for both input bins and output bins. 	}
	\label{fig:matrix_p-d_dn}
\end{figure}

\begin{figure}
	\centering
	\includegraphics[scale=0.21]{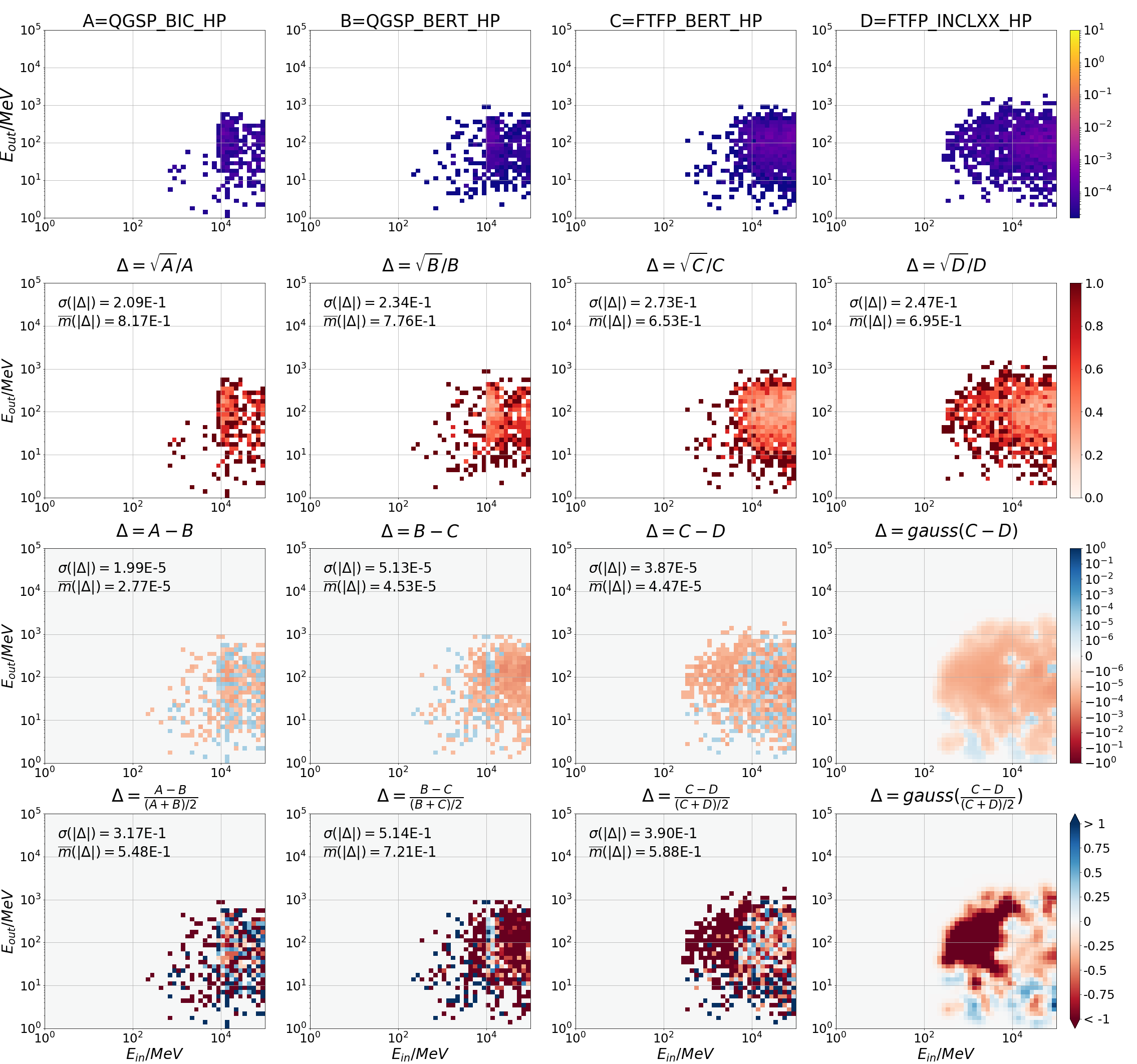}
	\caption{Matrices which describe the surface downward tritons created by the primary protons. First row: the normalized Martian atmospheric matrices of primary protons generating surface downward directed $^3$H ions from physics lists A, B, C and D (Table \ref{table:models}). 
	Second row: Normalized statistical uncertainty of the matrices. 
	Third row: Image Differencing Matrices (IDM) of model A and B (1st column),  B and C (2nd column), C and D (3rd column) which is also visualized with a 2D Gaussian filter applied (4th column). 
	Fourth row: the normalized IDM. 
	$\bar{m}$ represents the average value in each matrix (while bins with zero statistics have been excluded)  and $\sigma$ is the standard deviation.
	All matrices are shown in the energy range from 1 MeV to 100 GeV (in logarithmic scale) for both input bins and output bins. 	}
	\label{fig:matrix_p-t_dn}
\end{figure}

\begin{figure}
	\centering
	\includegraphics[scale=0.21]{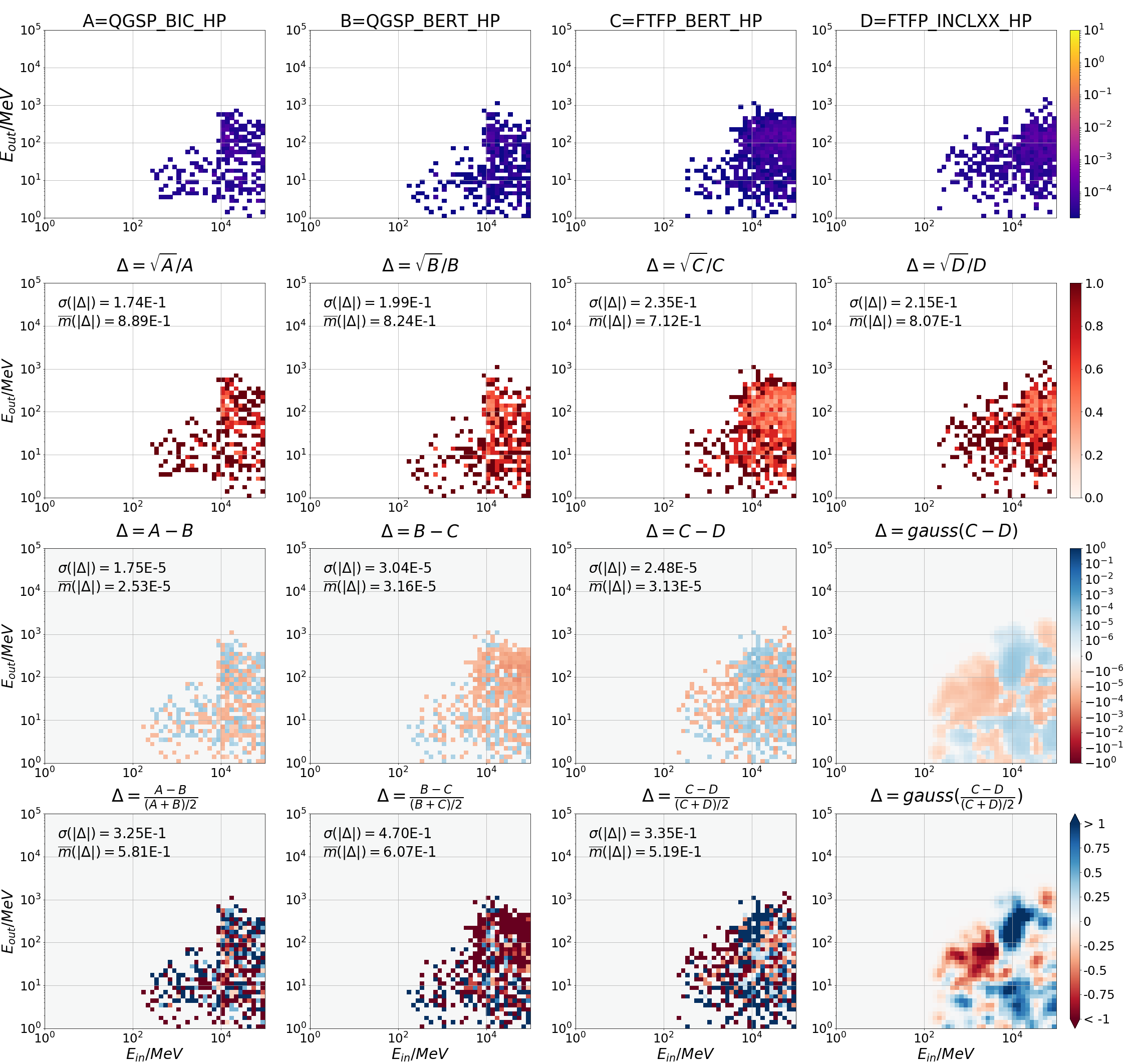}
	\caption{Matrices which describe the surface downward $^4$He particles created by the primary protons. First row: the normalized Martian atmospheric matrices of primary protons generating surface downward directed $^4$He ions from physics lists A, B, C and D (Table \ref{table:models}). 
	Second row: Normalized statistical uncertainty of the matrices. 
	Third row: Image Differencing Matrices (IDM) of model A and B (1st column),  B and C (2nd column), C and D (3rd column) which is also visualized with a 2D Gaussian filter applied (4th column). 
	Fourth row: the normalized IDM. 
	$\bar{m}$ represents the average value in each matrix (while bins with zero statistics have been excluded)  and $\sigma$ is the standard deviation.
	All matrices are shown in the energy range from 1 MeV to 100 GeV (in logarithmic scale) for both input bins and output bins. 	}
	\label{fig:matrix_p-a_dn}
\end{figure}

\begin{figure}
	\centering
	\includegraphics[scale=0.21]{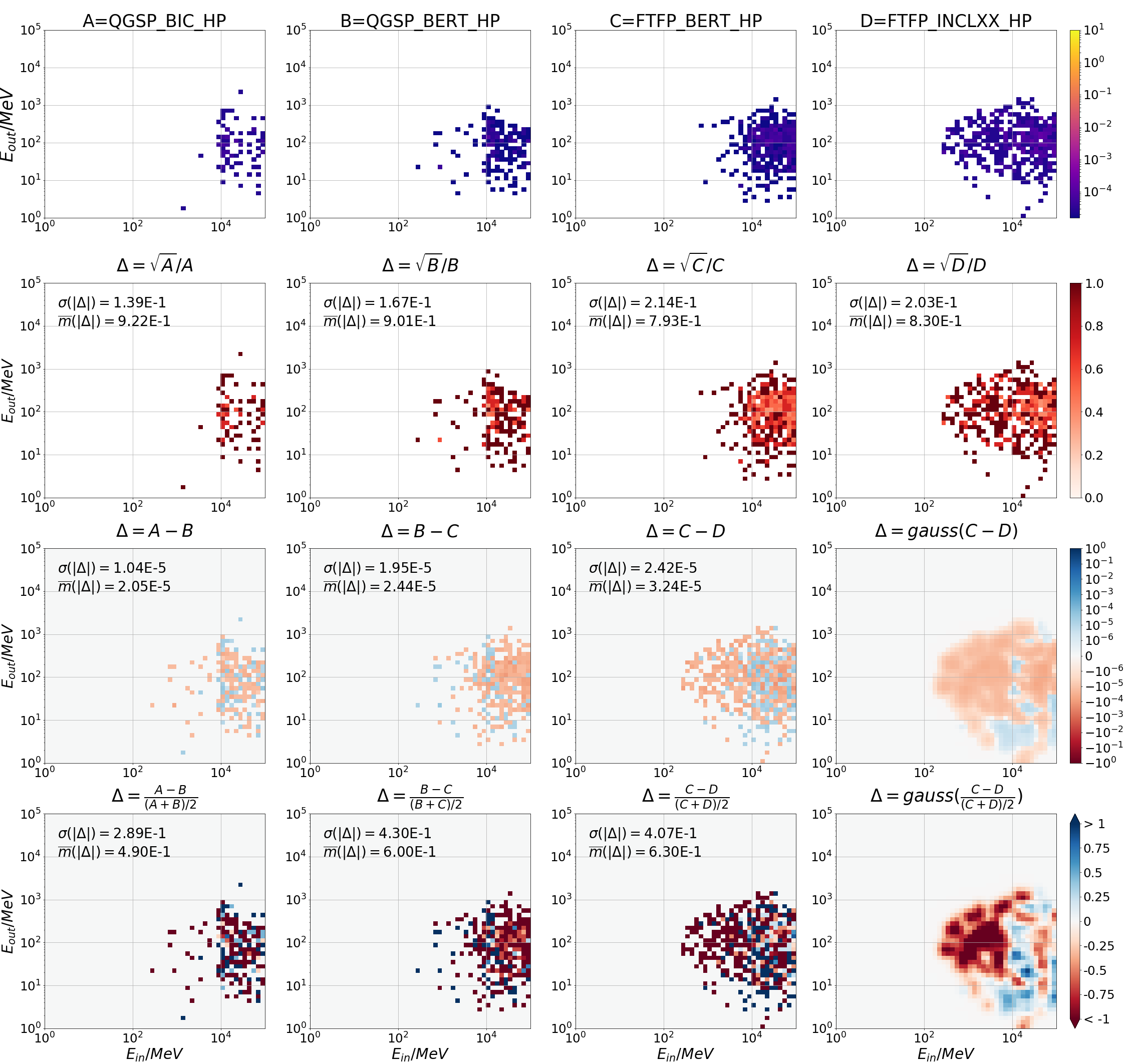}
	\caption{Matrices which describe the surface downward $^3$He particles created by the primary protons. First row: the normalized Martian atmospheric matrices of primary protons generating surface downward directed $^3$He ions from physics lists A, B, C and D (Table \ref{table:models}). 
	Second row: Normalized statistical uncertainty of the matrices. 
	Third row: Image Differencing Matrices (IDM) of model A and B (1st column),  B and C (2nd column), C and D (3rd column) which is also visualized with a 2D Gaussian filter applied (4th column). 
	Fourth row: the normalized IDM. 
	$\bar{m}$ represents the average value in each matrix (while bins with zero statistics have been excluded)  and $\sigma$ is the standard deviation.
	All matrices are shown in the energy range from 1 MeV to 100 GeV (in logarithmic scale) for both input bins and output bins. 	}
	\label{fig:matrix_p-h3_dn}
\end{figure}

\begin{figure}
	\centering
	\includegraphics[scale=0.21]{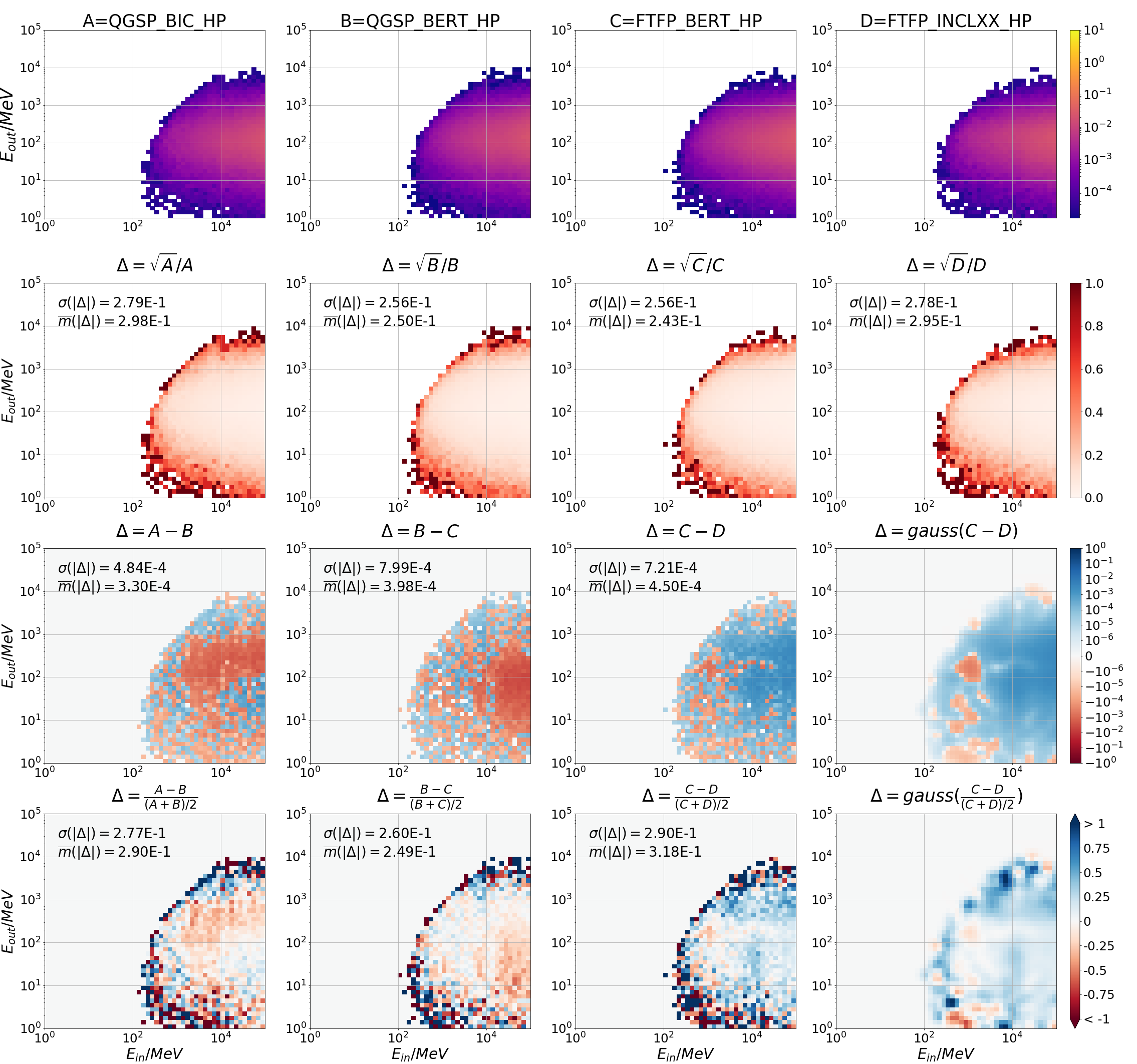}
	\caption{Matrices which describe the surface upward protons created by the primary protons. First row: the normalized Martian atmospheric matrices of primary protons generating surface upward directed protons from physics lists A, B, C and D (Table \ref{table:models}). 
	Second row: Normalized statistical uncertainty of the matrices. 
	Third row: Image Differencing Matrices (IDM) of model A and B (1st column),  B and C (2nd column), C and D (3rd column) which is also visualized with a 2D Gaussian filter applied (4th column). 
	Fourth row: the normalized IDM. 
	$\bar{m}$ represents the average value in each matrix (while bins with zero statistics have been excluded)  and $\sigma$ is the standard deviation.
	All matrices are shown in the energy range from 1 MeV to 100 GeV (in logarithmic scale) for both input bins and output bins. 	}
	\label{fig:matrix_p-p_up}
\end{figure}

Figure \ref{fig:matrix_p-p_dn} shows the matrices which describe the surfaced downward secondary protons which result from the primary protons under an atmospheric configuration as described in Section \ref{sec:model_mcd}. 
The general statistics of the matrices constructed from four different physics lists are rather similar as shown in the top row of the figure. 
We have also accounted for the Poisson uncertainty in the Monte Carlo simulations in each bin of the matrix and scaled it with the statistics in the corresponding bin to obtain the normalized uncertainties as plotted in the second row of the figure.
As shown, the normalized uncertainties are relatively small (below 20\%) in most bins apart from the low energy output bins where the bin widths are smaller (energies are binned into logarithmic scale) with lower statistics. 

To better quantify the differences between models, we have calculated the (relative) image difference between matrices of different models as shown in the last two rows of Figure \ref{fig:matrix_p-p_dn}, which we define as Image Differencing Matrices (IDM). 
It is visible that there are some systematic differences between different models and boundaries of patterns are formed at energies where physics processes are switching in the model. 
For instance, A-B IDM reflects the difference between BIC and BERT models for incoming particles below $\sim$ 10 GeV and the consistency between two physics lists at higher energies where QGSP is used for both models. 
Alternatively, B-C IDM shows the difference between QGSP and FTFP at high energies and the agreement of BERT at lower energies.  
In comparison, C-D shows a good agreement between FTFP\_BERT\_HP and  FTFP\_INCLXX\_HP apart from low output energy ranges where BERT seems to have a higher efficiency.  
This is because when simulating p-p interactions at energies below 20 MeV, FTFP\_INCLXX\_HP uses the models from the HP library. INCL is turned on only when one of the involved particles has the atomic number larger than one.

As shown in the bottom row, the mean relative differences are about 20.9\% between A and B, 23.2 \% between B and C and 17.7\% between C and D which are not much larger in comparison to the mean relative uncertainties of each model (second row).
Therefore the relative IDM indicates a general good agreement between different physics lists for protons generating secondary protons on the surface of Mars.
Besides, a 2D Gaussian filter has also been applied to the IDM of models C and D, shown in the last column of the last two rows, in order to smooth out features resulting from fine bins with low statistics. 

Figures \ref{fig:matrix_p-d_dn} and \ref{fig:matrix_p-t_dn} show the matrices of primary protons generating secondary deuterium and tritium hydrogen isotopes detected as downward particles on the surface of Mars. 
Compared to the generation of secondary protons, the p-d and p-t reaction probability is much lower as can be seen in the matrices of the top panels which have the same color bar scale as the proton-proton matrices.
The deuterium and tritium matrices lack the high energy diagonal component, while most secondaries are due to spallation reactions of high energy protons (above a few GeV) with the atmosphere nucleus.
Model D is clearly more efficient in generating deuterium and tritium particles via such reactions as is readily seen in the model D matrices.
Model A in the range with the binary cascade model is least efficient in such productions. 
Indicated by the IDM between different models, in the energy range below $\sim$ 10 GeV, deuteron and triton generation efficiency of different models follow: BIC $<$ BERT $<$ INCL; at higher energies, FTFP has a slightly higher efficiency than QGSP.  

Figure \ref{fig:matrix_p-a_dn} shows the matrices of primary protons resulting into surface downward $^4$He ions. 
These reactions are even more rare than the above proton-deuteron and proton-triton cases. 
Despite of the low statistics of such reactions, the IDM between A and B indicates a slightly higher efficiency of BERT and the IDM between B and C reflects a higher production rate in FTFP compared to QGSP. 
And it is also noticeable from the IDM between C and D that INCL predicts a higher probability of such spallation interactions in the relevant energy range. 

Finally, Figure \ref{fig:matrix_p-h3_dn} shows surface downward $^3$He ions induced by primary protons. 
The features of matrices in this case are very similar to those of deuterium, tritium and $^4$He ions in Figures \ref{fig:matrix_p-d_dn}, \ref{fig:matrix_p-t_dn}, and \ref{fig:matrix_p-a_dn} indicating similar cascading processes in generating these different types of secondaries. 
However $^3$He ions have a slightly lower statistics than $^4$He ions which may suggest some of the $^3$He ions are from fragmented $^4$He ions first generated in the atmosphere. 

Primary protons and their secondaries generated in the atmosphere may reach the Martian surface and generate upward particles in the regolith.
As an example, we only show the matrices constructed for the case of upward protons as shown in Figure \ref{fig:matrix_p-p_up}.
Similar to the downward proton case in Figure \ref{fig:matrix_p-p_dn}, the IDM of A-B and B-C show higher efficiency of BERT in the middle energy range and FTFP in the high energy part. 
The IDM of C-D shows features that are more similar to those of downward protons rather than those of deuterium, tritium and $^3$He particles, indicating that these are mostly downward propagating protons reaching the soil and get re-directed upwards rather than protons produced via spallation in the regolith. 
This is also suggested by the similar atmospheric cutoff energies of primary protons when comparing the downward and upward secondary proton matrices. 
The atmospheric cutoff energy is about 160 and 200 MeV in Figures \ref{fig:matrix_p-p_dn} and \ref{fig:matrix_p-p_up} respectively which approximately corresponds to the minimum energy a proton needs to traverse through $\sim$ 22 g/cm$^2$ of atmosphere. 

\subsubsection{Primary $^4$He ion matrices}
\begin{figure}
	\centering
	\includegraphics[scale=0.21]{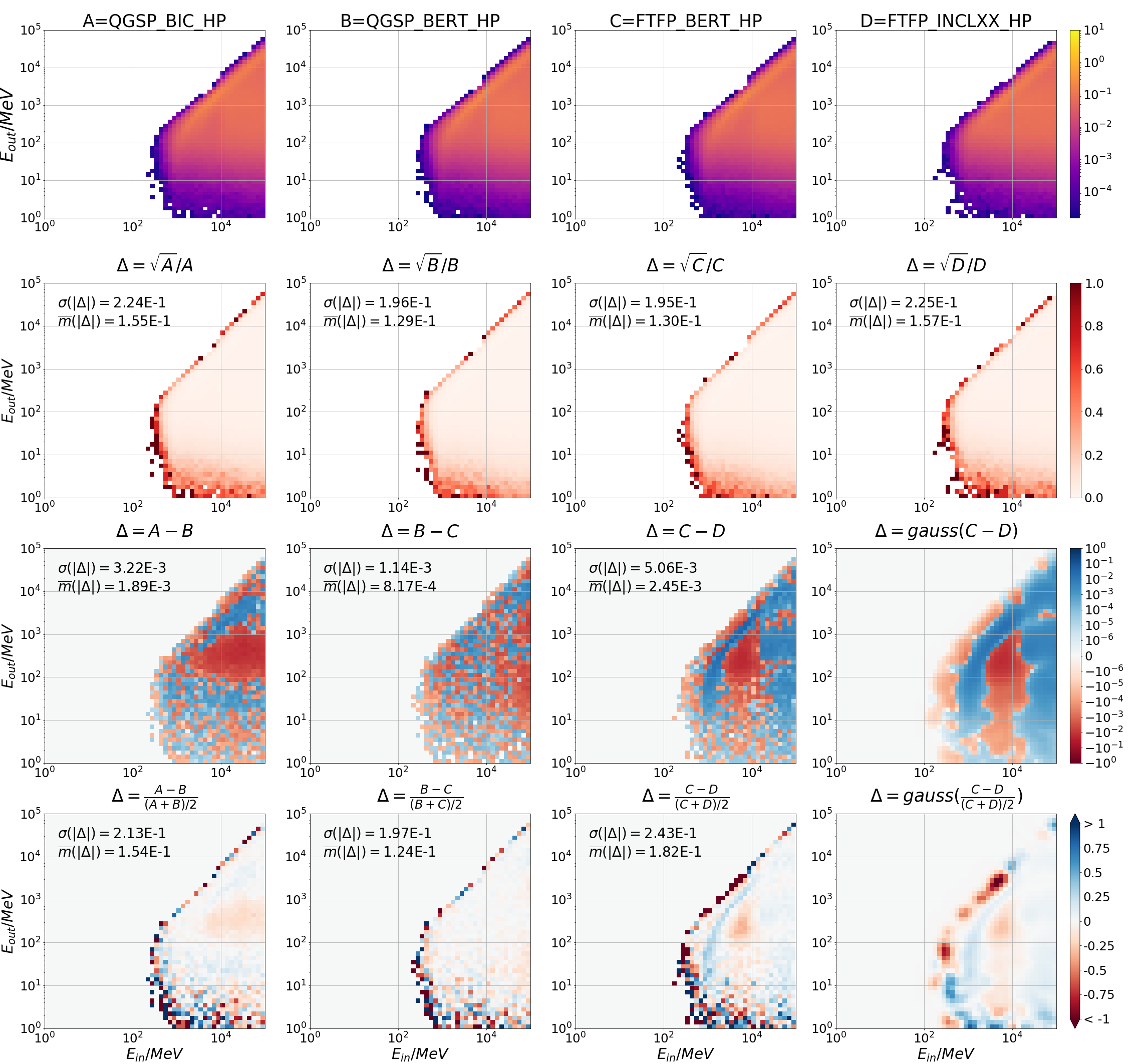}
	\caption{Matrices which describe the surface downward protons created by the primary $^4$He particles. 
	First row: the normalized Martian atmospheric matrices of primary $^4$He generating surface downward directed protons from physics lists A, B, C and D (Table \ref{table:models}). 
	Second row: Normalized statistical uncertainty of the matrices. 
	Third row: Image Differencing Matrices (IDM) of model A and B (1st column),  B and C (2nd column), C and D (3rd column) which is also visualized with a 2D Gaussian filter applied (4th column). 
	Fourth row: the normalized IDM. 
	$\bar{m}$ represents the average value in each matrix (while bins with zero statistics have been excluded)  and $\sigma$ is the standard deviation.
	All matrices are shown in the energy range from 1 MeV to 100 GeV (in logarithmic scale) for both input bins and output bins. 	} 
	\label{fig:matrix_a-p_dn}
\end{figure}

\begin{figure}
	\centering
	\includegraphics[scale=0.21]{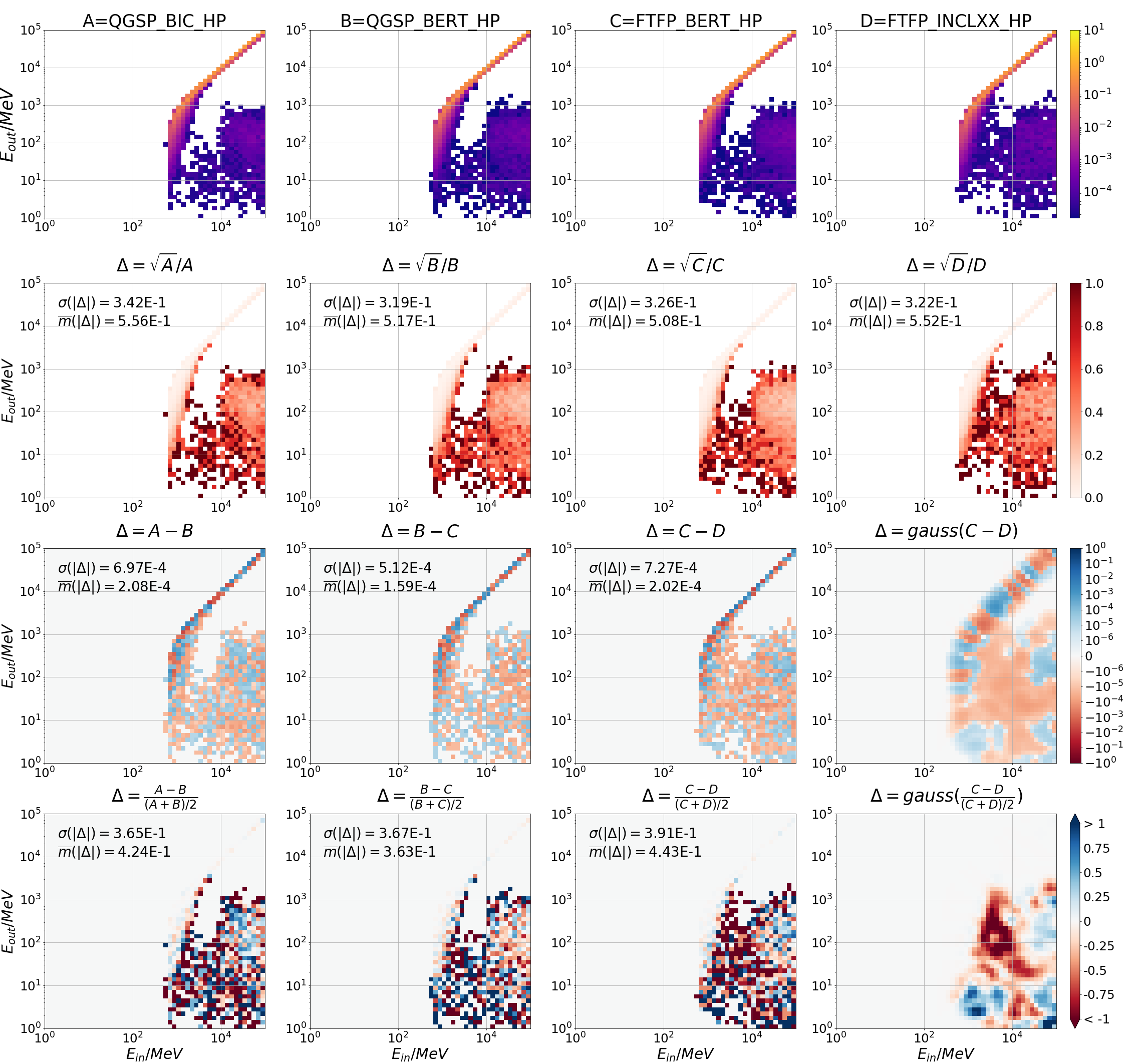}
	\caption{Matrices which describe the surface downward $^4$He created by the primary $^4$He particles. 
	First row: the normalized Martian atmospheric matrices of primary $^4$He generating surface downward directed $^4$He from physics lists A, B, C and D (Table \ref{table:models}). 
	Second row: Normalized statistical uncertainty of the matrices. 
	Third row: Image Differencing Matrices (IDM) of model A and B (1st column),  B and C (2nd column), C and D (3rd column) which is also visualized with a 2D Gaussian filter applied (4th column). 
	Fourth row: the normalized IDM. 
	$\bar{m}$ represents the average value in each matrix (while bins with zero statistics have been excluded)  and $\sigma$ is the standard deviation.
	All matrices are shown in the energy range from 1 MeV to 100 GeV (in logarithmic scale) for both input bins and output bins. 	} 
	\label{fig:matrix_a-a_dn}
\end{figure}

\begin{figure}
	\centering
	\includegraphics[scale=0.21]{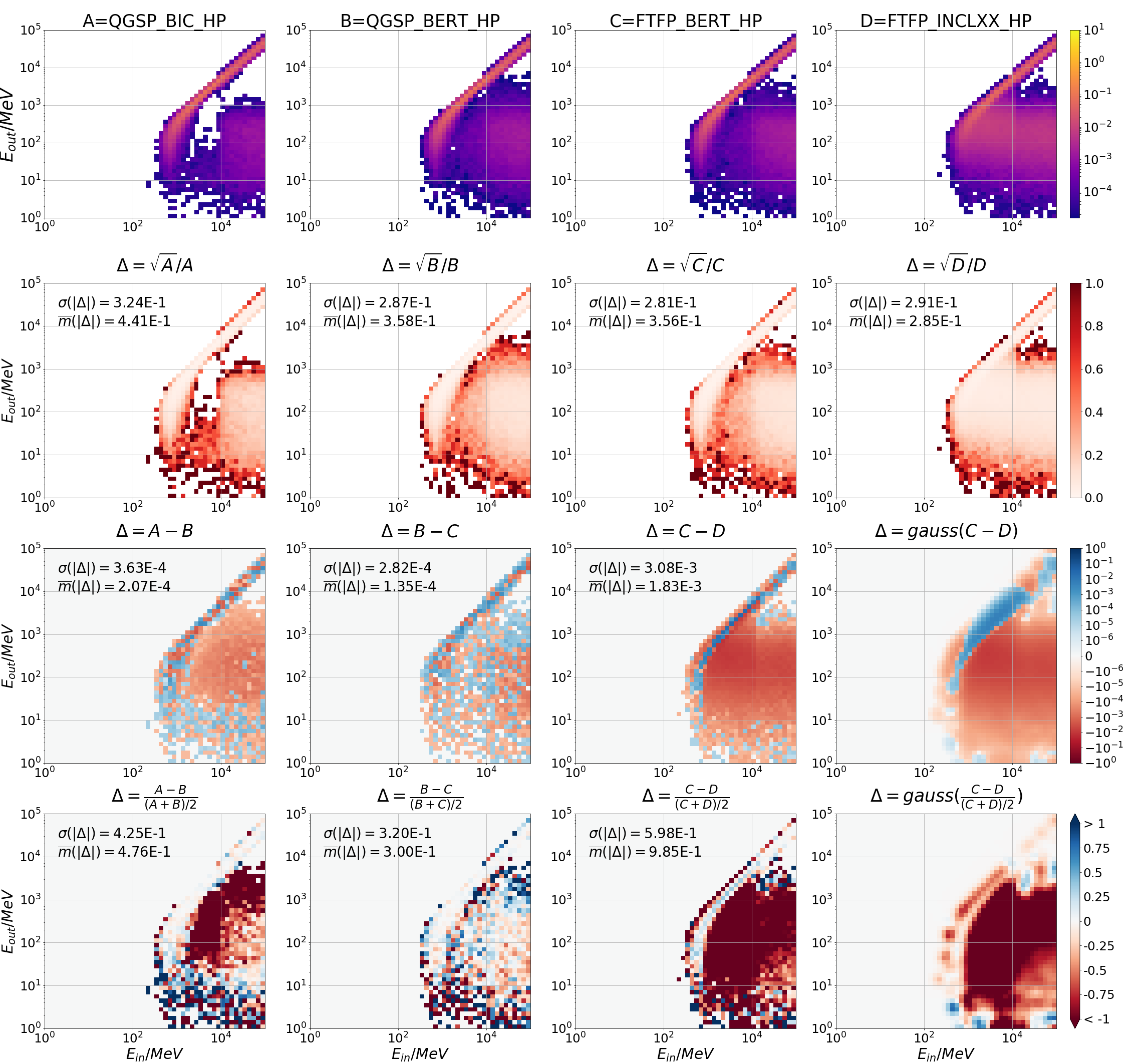}
	\caption{Matrices which describe the surface downward deuterons created by the primary $^4$He particles. 
	First row: the normalized Martian atmospheric matrices of primary $^4$He generating surface downward directed $^2$H from physics lists A, B, C and D (Table \ref{table:models}). 
	Second row: Normalized statistical uncertainty of the matrices. 
	Third row: Image Differencing Matrices (IDM) of model A and B (1st column),  B and C (2nd column), C and D (3rd column) which is also visualized with a 2D Gaussian filter applied (4th column). 
	Fourth row: the normalized IDM. 
	$\bar{m}$ represents the average value in each matrix (while bins with zero statistics have been excluded)  and $\sigma$ is the standard deviation.
	All matrices are shown in the energy range from 1 MeV to 100 GeV (in logarithmic scale) for both input bins and output bins. 	} 
	\label{fig:matrix_a-d_dn}
\end{figure}

\begin{figure}
	\centering
	\includegraphics[scale=0.21]{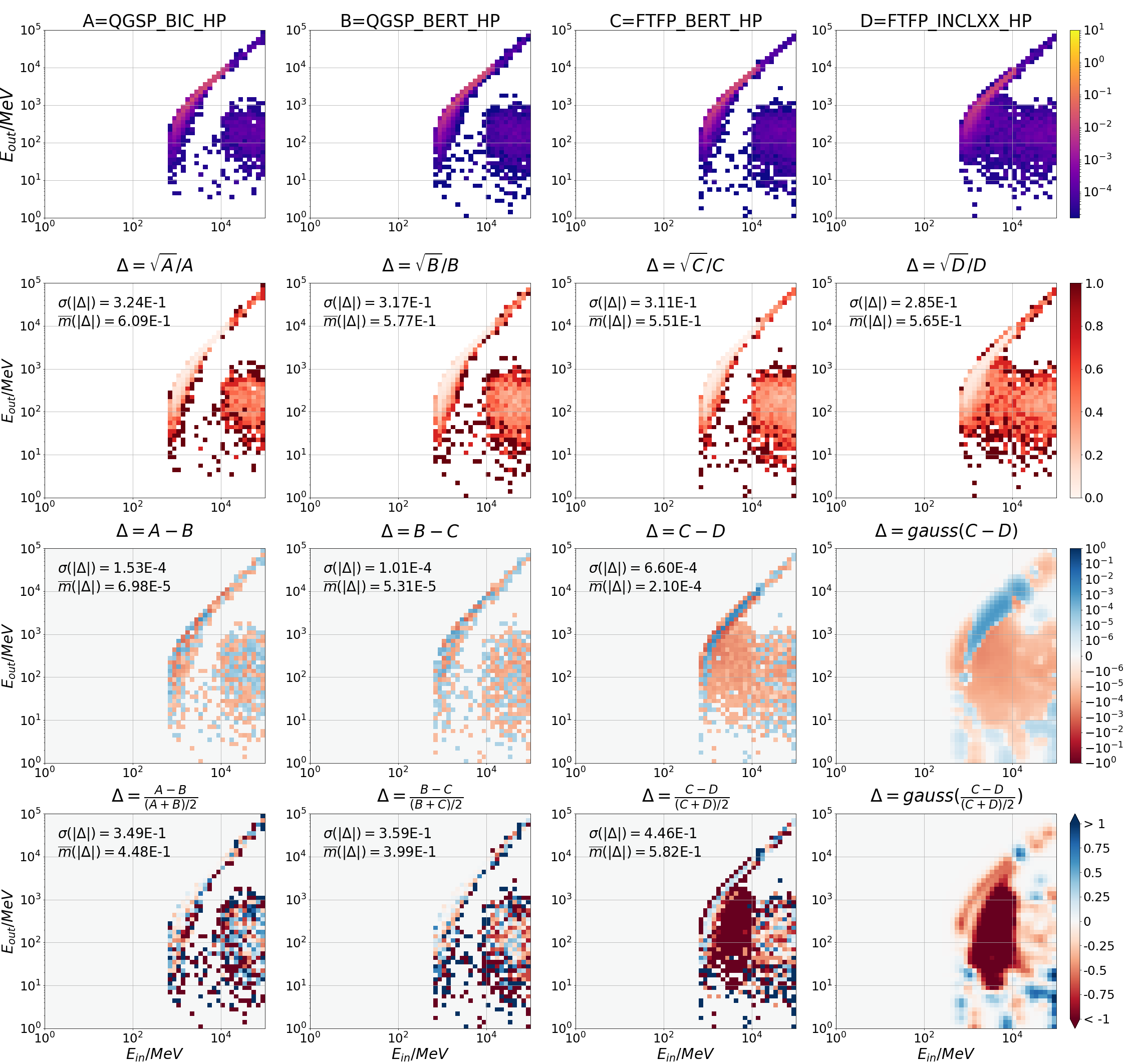}
	\caption{Matrices which describe the surface downward $^3$He created by the primary $^4$He particles. 
	First row: the normalized Martian atmospheric matrices of primary $^4$He generating surface downward directed $^3$He from physics lists A, B, C and D (Table \ref{table:models}). 
	Second row: Normalized statistical uncertainty of the matrices. 
	Third row: Image Differencing Matrices (IDM) of model A and B (1st column),  B and C (2nd column), C and D (3rd column) which is also visualized with a 2D Gaussian filter applied (4th column). 
	Fourth row: the normalized IDM. 
	$\bar{m}$ represents the average value in each matrix (while bins with zero statistics have been excluded)  and $\sigma$ is the standard deviation.
	All matrices are shown in the energy range from 1 MeV to 100 GeV (in logarithmic scale) for both input bins and output bins. 	} 
	\label{fig:matrix_a-h3_dn}
\end{figure}

Similar to the atmospheric response matrices of primary protons, the primary $^4$He ion interaction with the Martian atmosphere and the generation of secondaries have been  mapped to matrices with some examples shown in Figures \ref{fig:matrix_a-p_dn}, \ref{fig:matrix_a-a_dn}, \ref{fig:matrix_a-d_dn} and \ref{fig:matrix_a-h3_dn}. 

Figure \ref{fig:matrix_a-p_dn} displays the matrices for $^4$He generated surface protons. The general shape from four physics lists looks rather similar and also very comparable to the proton-proton matrices in Figure \ref{fig:matrix_p-p_dn}, indicating that a substantial amount of $^4$He ions fragment into protons in the atmosphere and follow the same energy loss processes as in the proton-proton case. 
It is noticeable that model D predicts a slightly higher efficiency in the primary energy range from a few GeV to $\sim$ 20 GeV corresponding to the INCL model in the physics list of model D. 

Figure \ref{fig:matrix_a-a_dn} shows the $^4$He-$^4$He matrices from four physics lists which have similar structures. It is visible that there are 3 components in these matrices. 
The first is represented by the top-right diagonal component showing some $^4$He particles with energies $\ge \sim$ 2-3 GeV arriving at the Martian surface as $^4$He ions with similar energies, i.e, these {relativistic} particles propagate through the 22 g/cm$^2$ of atmosphere without any interactions. 
The second component is shown as a vertical branch connecting to the lower-left part of the diagonal structure. These are primary ions with energies between $\sim$ 650 MeV and $\sim$ 3-4 GeV which traverse through the atmosphere with substantial ionization energy loss. 
The third component is the blob structure at the lower-right part of the matrices (similar to the p-d, p-t matrices in Figures \ref{fig:matrix_p-d_dn} and \ref{fig:matrix_p-t_dn}) which represent high energy primary particles interacting with the atmosphere and generating secondary $^4$He ions via spallation process. 
There is a gap beneath the diagonal component while there is an enhanced production of $^4$He produced protons at this range (Figure \ref{fig:matrix_a-p_dn}). This suggests that $^4$He particles easily fragment into protons as they traverse through the atmosphere. 

As shown by the IDM in the lower panel, all four models result in very small differences especially for the first component. 
The difference between different models is mostly due to statistics as the IDM shows rather similar features compared to the uncertainty matrices and the mean values of IDM are comparable to the mean matrix uncertainties.
Model D seems to predict a larger effect of the third component, especially in the energy range where the INCL model is applied.

Figure \ref{fig:matrix_a-d_dn} represents the $^4$He-deuteron case and also shows the three-component structure in four models. 
Here the first and second components correspond to $^4$He ions fragmenting into $^2$H particles as they propagate through the atmosphere. 
It is also shown by the IDM of A-B that Bertini cascade is more efficient and Binary cascade. 
The third component predicted by model D is much more enhanced compared to model A and B and the transition between the second and the third components in model D is much smoother. 
The average difference between model D and C is significant and as large as 100\%. 
Similar features of the matrices are also shown in Figure \ref{fig:matrix_a-h3_dn} for the $^4$He-$^3$He case with model D predicating an enhancement of $^3$He production especially at the range between the second and third components. 
This might be due to $^4$He being more efficient in fragmenting into $^3$He particles. 

\subsection{Comparison of Modeled spectra with RAD proton measurements}\label{sec:validation_RAD}

\begin{figure}[ht!]
\centering
\begin{tabular}{c}
\subfloat[input: GCR H + $^4$He, output:  downward H isotopes] { \includegraphics[trim=0 20 0 50,clip, scale=0.4]{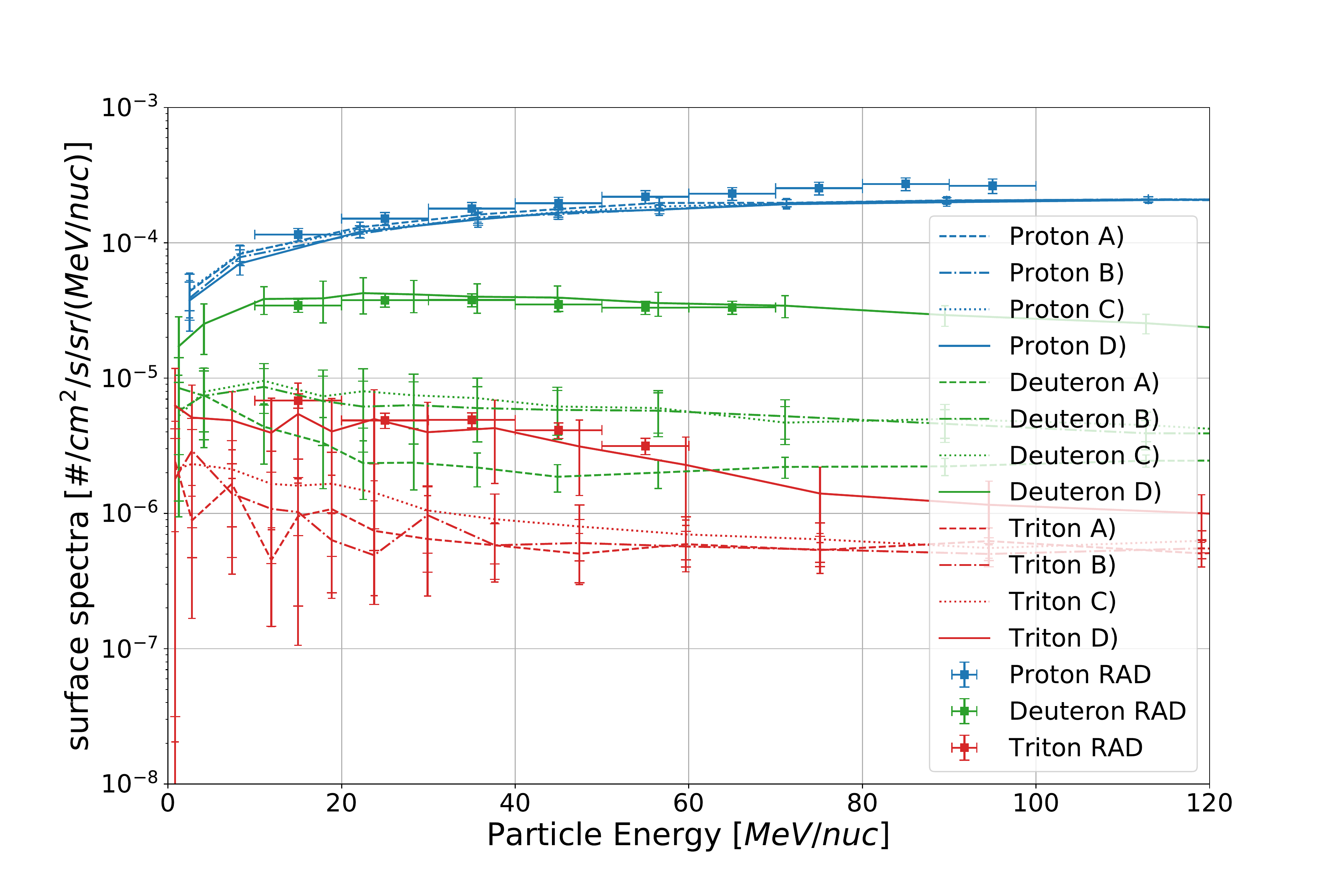}}\\ 
\subfloat[input: GCR H + $^4$He, output:  downward He isotopes] { \includegraphics[trim=0 20 0 50,clip, scale=0.4]{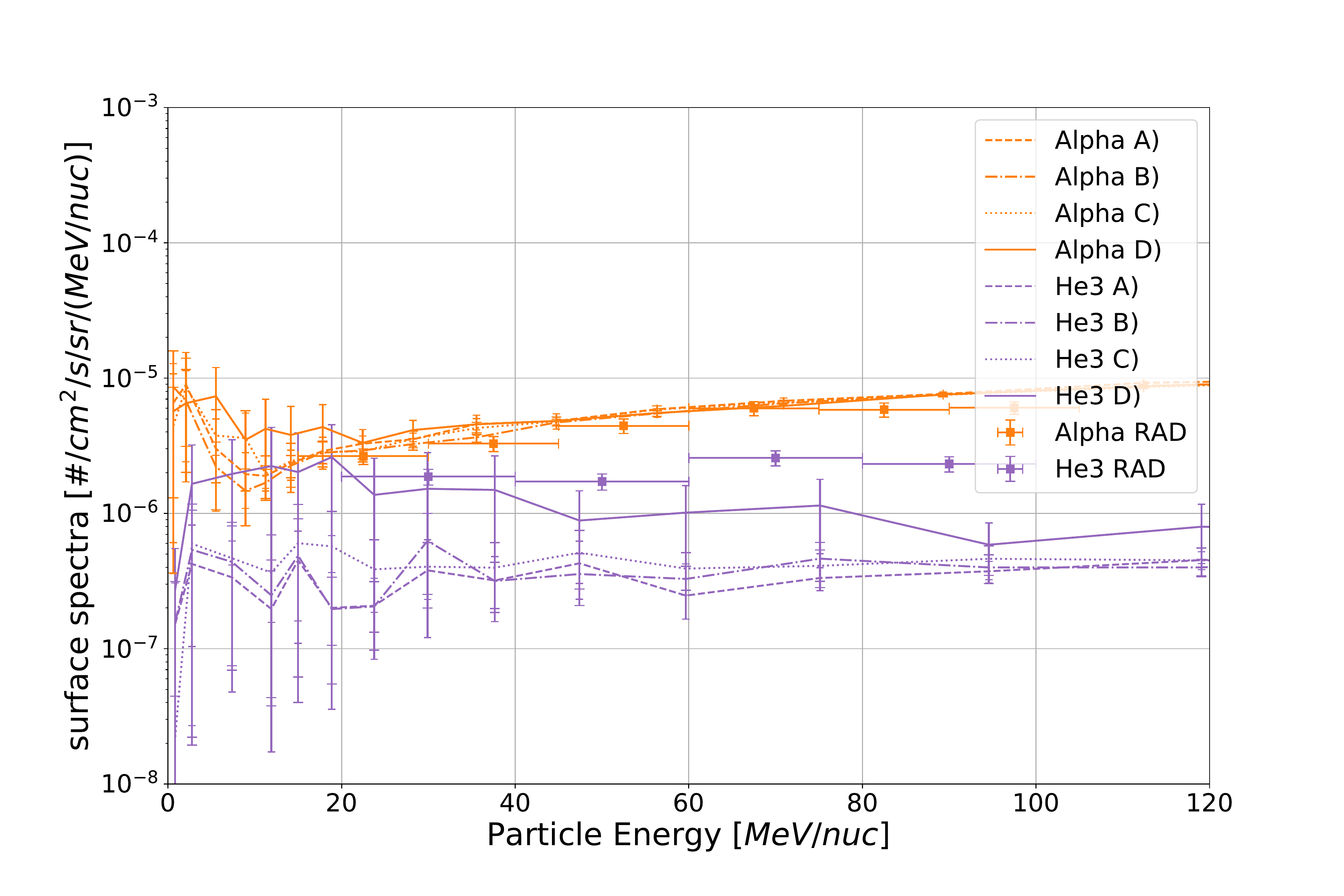}} 
\end{tabular}
\caption{Martian surface spectra of H (panel a) and He (panel b) isotopes induced by primary GCR proton and $^4$He particles modeled via four different physics lists (Table \ref{table:models}). The spectra {are} averaged within 36$^\circ$ degrees of downward zenith angle corresponding to the RAD view cone.}\label{fig:spec_models_rad}
\end{figure}

In order to generate surface secondary particle spectra, we fold the above matrices with the primary proton and helium GCR spectra calculated from a standard GCR model -- the Badhwar-O'Neill 2010 \citep[BON10,][]{badhwar2010}. 
The BON10 spectra have been computed using the spherically symmetric Fokker-Planck equation, with the Local Interstellar Spectrum (LIS) at the boundary of the heliosphere ($\sim$ 100 AU) as a boundary condition. 
The model uses an input parameter of the solar modulation $\Phi$, in units of volt, as derived from the International Sunspot Number. 
The modulation parameter $\Phi$ is a depiction of solar activity and its modulation magnitude on the GCR particle fluxes. 
The modeled spectra {at Earth} have been fitted and adjusted to measurements obtained by a multitude of instruments, including all available observations of GCR particle energy spectra from 1955 to 2010.
{Note that we use the near-Earth GCR spectra to approximate those at Mars with a distance of $\sim$ 1.5 AU to the Sun. The radial gradient of galactic cosmic rays in the inner heliosphere has been estimated to be about 3\% per AU for protons with energies around 1 GeV  \citep[e.g.,][]{gieseler2016spatial}. This translates into about 1.5\% of increase of GCR flux from Earth and Mars. However, this correction is much smaller compared to other systematic and statistical uncertainties in the models employed in the current work and therefore we omit the radial gradient of the heliospheric modulation of GCRs. }

The solar modulation parameter used here as input for the BON10 GCR model is equal to the average value, 550 MV, during the periods of the selected RAD measurements \citep{EHRESMANN20173}. 
Primary protons and $^4$He ions from the GCR model have been folded with the matrices to generate surface secondary particles. 
The resulting secondary spectra of protons, $^4$He and $^3$He ions, deuterons, tritons in the energy range of 10 to 100 MeV/nuc in the RAD view cone have been compared with MSL/RAD surface measurements as shown in Figure \ref{fig:spec_models_rad}. 
{Although RAD is not always positioned perpendicular to the surface as the rover body is inclined occasionally, \citet{wimmer2015} have found from the measurement that the variability of downward particle fluxes with the zenith angle is small. }
Therefore a separate set of matrices with surface particles constrained within the RAD view cone ($<$ 36$^\circ$ degrees of downward zenith angle) has been constructed.
Besides, we have also propagated the Poisson uncertainties from the simulations through the matrix multiplication for obtaining the uncertainties of the modeled surface spectra (Equation \ref{eq:matrix_all}). 

The proton spectra modeled by four different physics lists all agree very well with each other and also the RAD measurement. The modeled fluxes are slightly lower than the observed spectra which is reasonable as GCR particles heavier than $^4$He have not been considered in the model.
Despite the large uncertainties due to low statistics, it is clearly shown that the fluxes for both deuterium and tritium particles from model A B and C are considerably lower than the measured spectra while those predicted by model D match well with the data. 
As explained in Section \ref{sec:validation_matrix} and shown in Figures \ref{fig:matrix_p-d_dn}, \ref{fig:matrix_p-t_dn}, and \ref{fig:matrix_a-d_dn}, model D has a higher efficiency in generating secondary particles via spallation process (or the third component) which are located at the energy range RAD measures. 

For Helium isotopes, all four models show a good agreement of $^4$He ions, although with a slightly higher flux, compared to the RAD data. 
For $^3$He case, model D again has a better prediction in comparison to the RAD measurement in the energy range of 10 - 100 MeV/nuc. 
The higher production rate of $^3$He in model D is clearly shown in Figures \ref{fig:matrix_p-h3_dn} and \ref{fig:matrix_a-h3_dn} as explained in Section \ref{sec:validation_matrix}. 

\begin{sidewaysfigure}
\centering
\begin{tabular}{ccc}
\subfloat[RAD view cone H isotopes] { \includegraphics[trim=0 20 0 50,clip, scale=0.20]{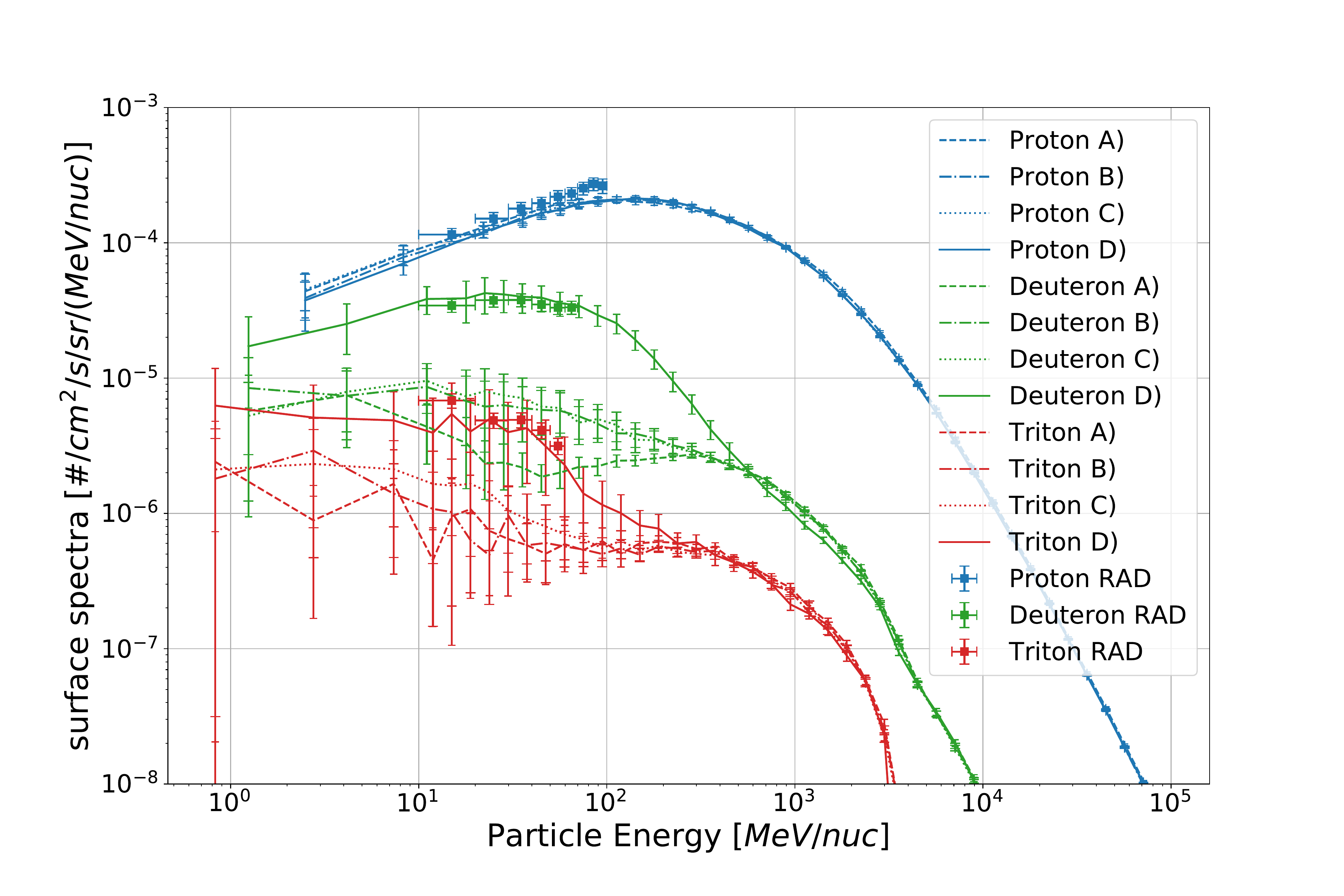}} &
\subfloat[Downward H isotopes] { \includegraphics[trim=0 20 0 50,clip, scale=0.20]{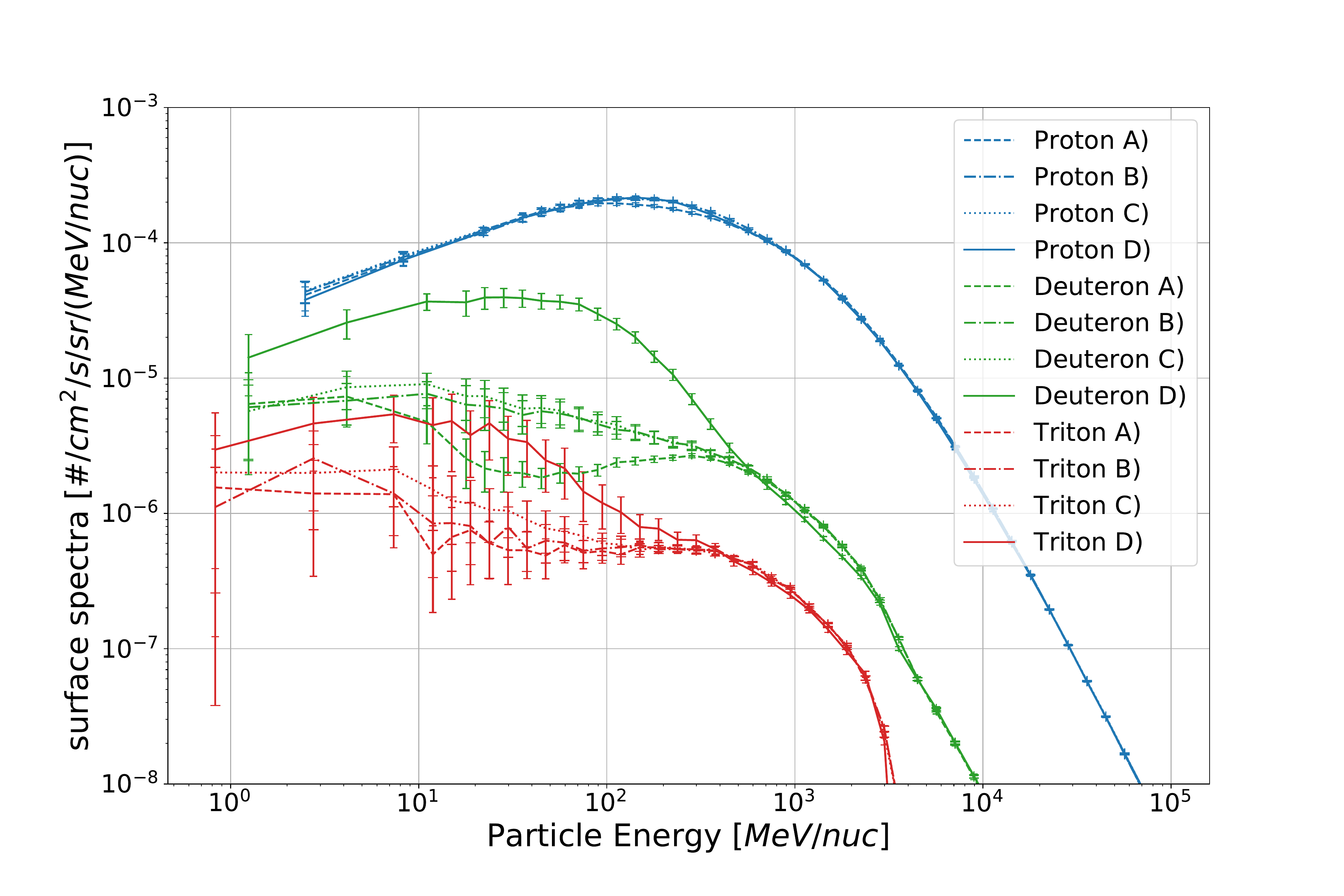}} &
\subfloat[Upward H isotopes] { \includegraphics[trim=0 20 0 50,clip, scale=0.20]{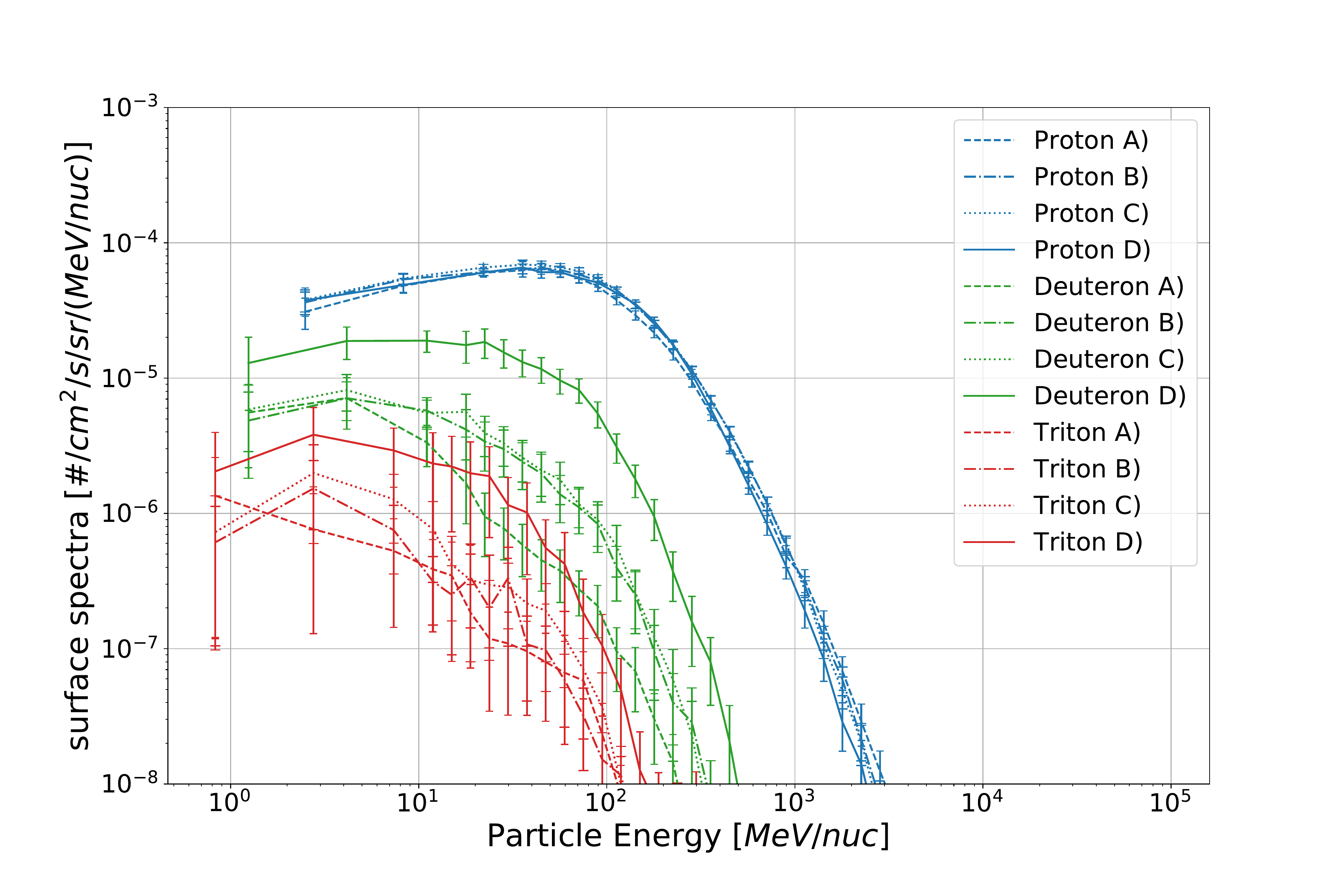}} \\
\subfloat[RAD view cone He isotopes] { \includegraphics[trim=0 20 0 50,clip, scale=0.20]{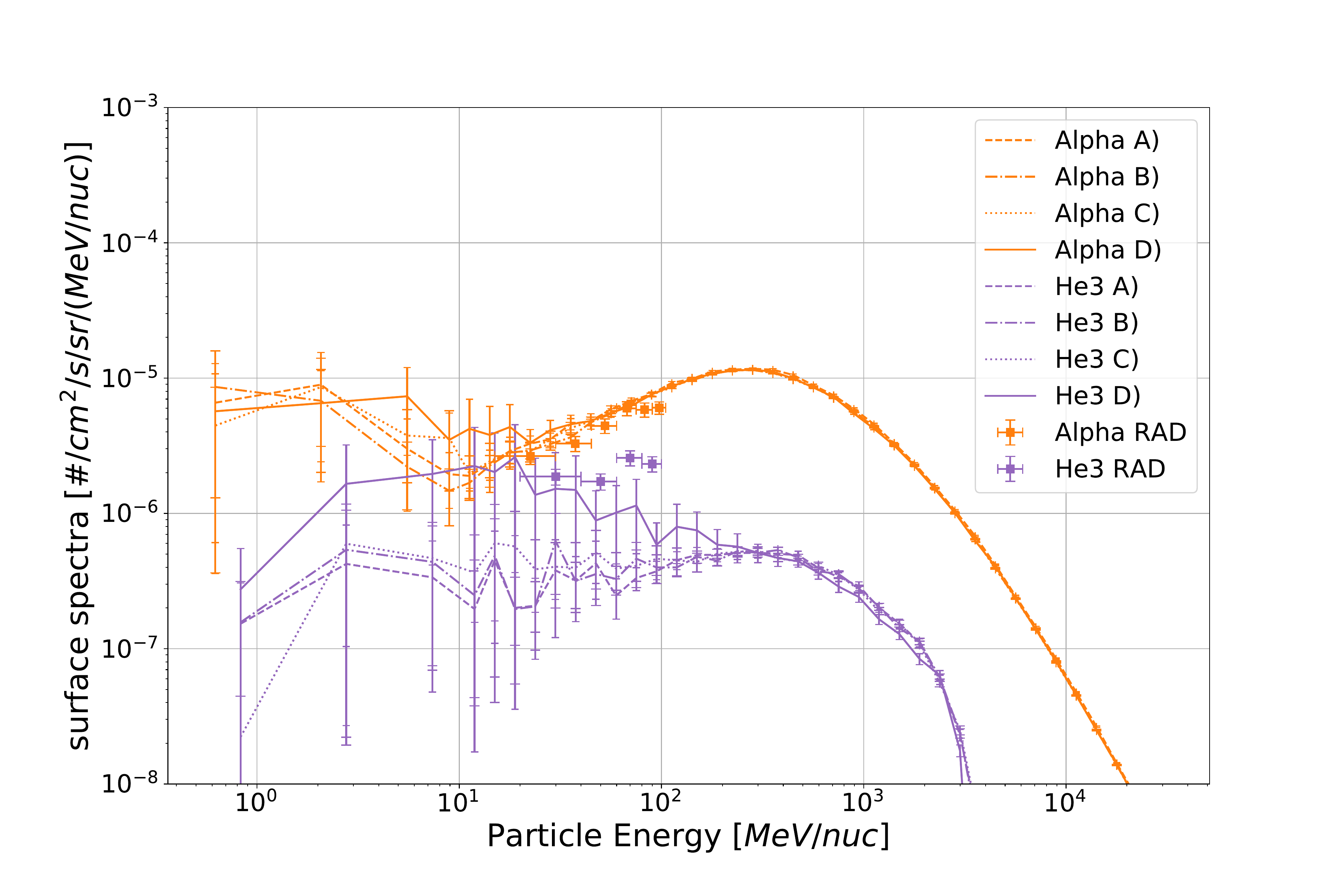}} &
\subfloat[Downward He isotopes] { \includegraphics[trim=0 20 0 50,clip, scale=0.20]{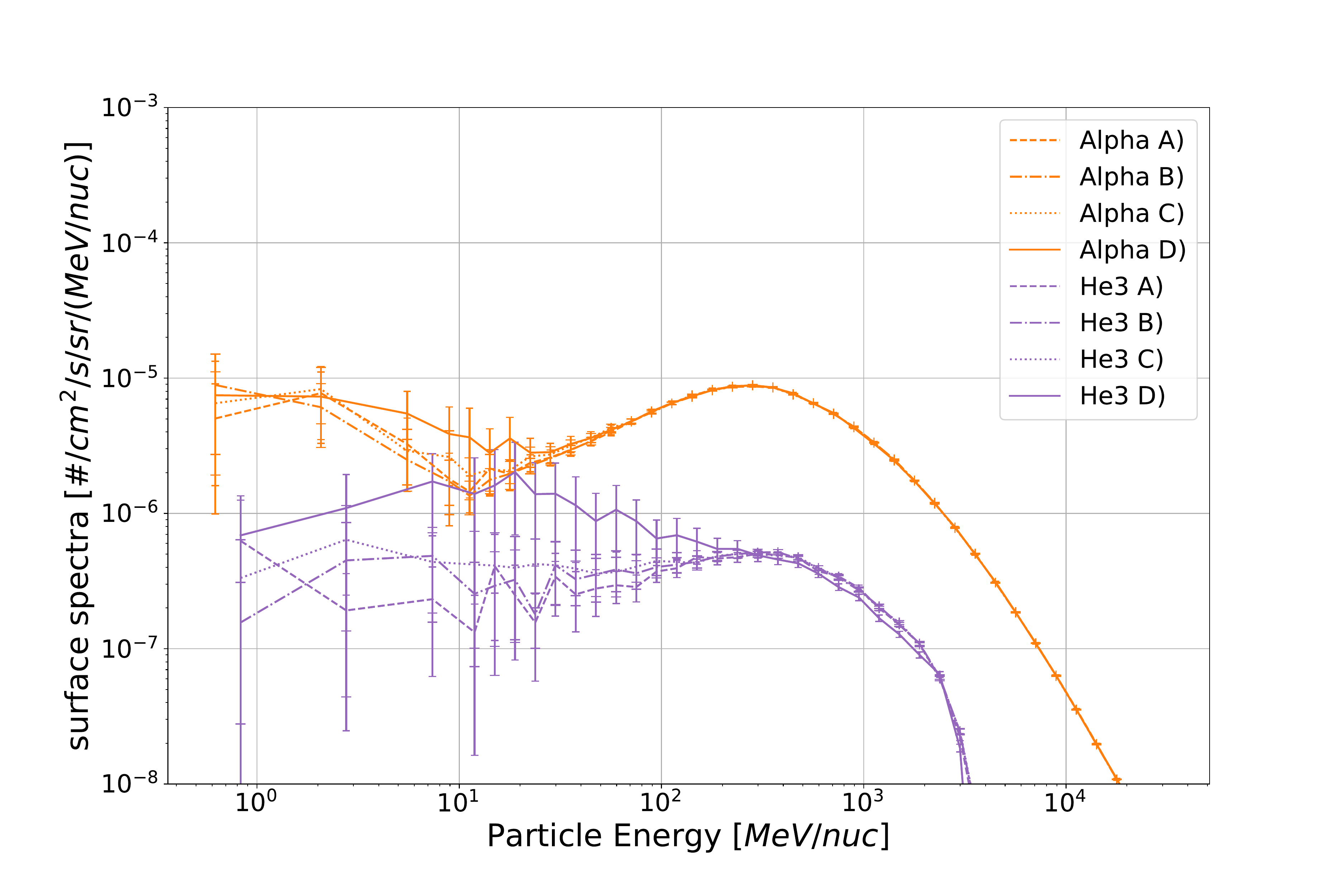}} &
\subfloat[Upward He isotopes] { \includegraphics[trim=0 20 0 50,clip, scale=0.20]{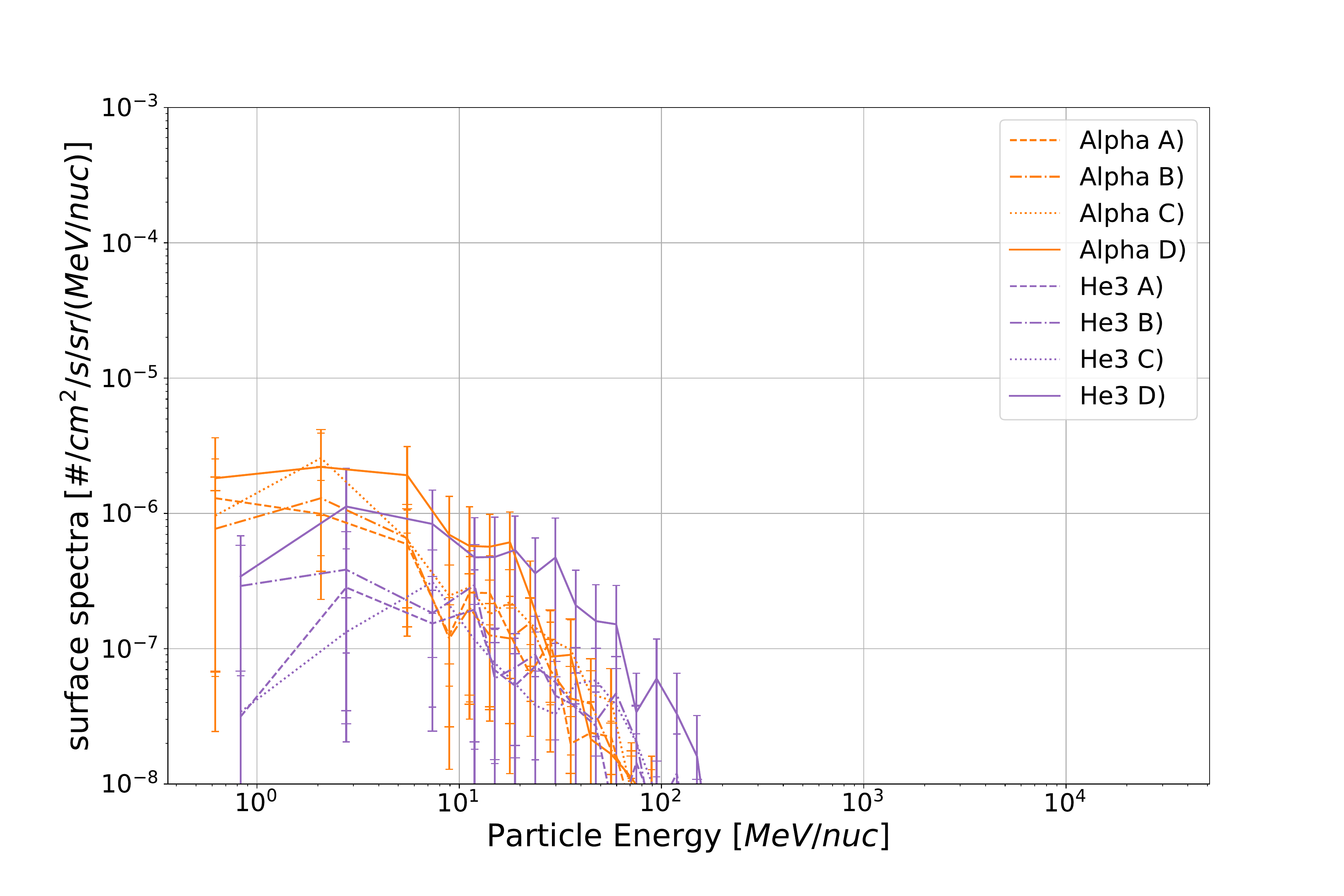}} \\
\end{tabular}
\caption{Martian surface spectra of H (panel a, b, c) and He (panel d, e, f) isotopes induced by primary GCR proton and $^4$He particles modeled via four different physics lists (Table \ref{table:models}). The spectra {are} averaged within different solid angles: a) and d) for the RAD view cone, b) and e) for the downward direction and d) for the upward direction. }\label{fig:spec_models_full}
\end{sidewaysfigure}

We have also compared the modeled spectra in a larger range beyond the energy range of the RAD data as shown in Figure \ref{fig:spec_models_full} a) and d) for H and He isotopes respectively. 
For protons and $^4$He ions, all models agree with each other at different energy ranges. 
However for deuterons, tritons and $^3$He ions, model D shows significant enhancement of the flux compared to other models for particles up to $\sim$ 500 MeV/nuc. 
At higher energies the discrepancy between different models is much smaller as also shown in the deuterium, tritium and $^3$He matrices. 
The RAD measurement locates within the energy range where predictions from different physics lists differ the most and the current comparison strongly supports the FTFP\_INCLXX\_HP model for GCR particles transported through the Martian environment. 

The angular dependence of the secondary particles on the surface of Mars is also very important for future human exploration of Mars and understanding radiation effects on Mars. The downward spectra within the RAD view cone shown in Figure \ref{fig:spec_models_full}(a) and those averaged over the whole downward-directed solid angle shown in panel (b) have very similar levels of flux. 
This is consistent with the results found by \citet{wimmer2015} from the RAD measurement that within 15$^\circ$ of the Rover tilt angles, the radiation field is mostly isotropic.

However the upward spectra are rather dissimilar from the downward spectra and all upward spectra lack the high energy component ($> \sim$ GeV) compared to downward spectra. 
Particle fluxes of protons and $^4$He ions are particularly lower than {in} the downward direction where more primary GCR particles contribute to the surface downward radiation.
As the upward flux, mostly produced in the Martian regolith, is much lower than the {downward} flux, this suggests the effectiveness of shielding using the Martian regolith for potential human habitat on Mars. 

\section{Summary and Conclusion}
We have applied the novel tool AtRIS \citep{banjac2018} with a specific implementation of the Martian atmospheric and regolith structure to model the radiation environment at Mars. 

We have validated the accuracy of the AtRIS model when applied with four different physics lists to the Martian environment, especially for the generation of charged particle spectra by primary protons and $^4$He ions. 
We first visualized the atmospheric response matrix of each primary-secondary chain via a two-dimensional histogram where the energy-dependent efficiency of primary particles generating secondaries is nicely shown. 
We compared these matrices obtained via four different physics lists of GEANT4 by calculating and visualizing their image-differencing matrices (IDM). 
The IDM results suggest that in general {the} Bertini cascade model is more efficient than the Binary cascade model while INCL model is most productive especially for deuterium, tritium and $^3$He particles. 
For two primary particle types of protons and $^4$ He ions, four different physics lists agree with each other reasonably well. 
For other hydrogen and helium isotopes, model D FTFP\_INCLXX\_HP generates significantly more secondaries due to spallation processes of primary particles in the energy range of $\sim$ GeV and 20 GeV where the Li\`ege Intra-nuclear Cascade model is applied, as shown in Figures \ref{fig:matrix_p-d_dn}, \ref{fig:matrix_p-t_dn}, \ref{fig:matrix_p-h3_dn}, \ref{fig:matrix_a-d_dn} and \ref{fig:matrix_a-h3_dn}.  

To benchmark the AtRIS model based on different GEANT4 physics lists when applied to the Martian environment, we compared the AtRIS predicted spectra with the energetic particle spectra recently measured by MSL/RAD on the surface of Mars \citep{EHRESMANN20173}. 
We folded the above atmospheric response matrices with BON GCR spectra of protons and $^4$He particles to obtain the secondary particle spectra of Hydrogen and Helium isotopes: protium, deuterium, tritium, $^3$He and $^4$He particles. 
This process has summed up the same secondary type generated by both proton and $^4$He primaries. But we have ignored the contribution by other heavier GCR particles. This may have caused the slightly smaller flux of proton in all four models in comparison with the MSL/RAD measurement. 
The comparison shows that all models have an agreed prediction of the proton and helium flux in comparison to the RAD measurement in the energy range of 10 - 100 MeV/nuc. 
However, model D has a higher prediction of deuterium, tritium and $^3$He flux and a better match with the RAD data. 
This is due to the same reason we have already observed in the atmospheric response matrices. 

In general, the good agreement between AtRIS and the actual measurement extends the validation of AtRIS to non-Earth systems. For the specific case of a very thin atmosphere, we have analyzed how different physics lists may impact the generation of secondary particles through the Martian atmosphere.
While we have seen that FTFP\_INCLXX\_HP provides a better agreement with the MSL/RAD data for the specific cases that were examined, a further study is needed to validate this for {heavier primary ions and other secondary particles as well as for} planets with thicker atmospheres and different atmospheric compositions (Earth-like and Venus-like). 
{For instance, secondary muons are a major surface radiation component at the largest moon of Saturn -- Titan. Heavier ions and neutrons have more enhanced biological effectiveness and should also be carefully examined and included when using AtRIS to predict the equivalent radiation dose.}
 
In order to make AtRIS easily accessible by the community, we are preparing an online documentation including examples for AtRIS in the form of a wiki page. In 2019, AtRIS will be published under a GNU GPL licence. 

\begin{acknowledgements}
      Part of this work was supported by the German \emph{Deutsche Forschungs Gemeinschaft, DFG} project 282759267.
      The work is also supported by DLR and DLR's Space Administration grant numbers 50QM0501, 50QM1201 and 50QM1701 to the Christian Albrechts University, Kiel. 
      We thank Bent Ehresmann for the helpful discussions and support. 
      J.G. and R.F.W.S. acknowledge the International Space Science Institute, which enabled part of the collaborations in this paper through the ISSI International Team 353 “Radiation Interactions at Planetary Bodies”.
      The authors and the editor thanks Rami Vainio and an anonymous referee for their assistance in evaluating this paper.
\end{acknowledgements}



\end{document}